\documentclass[onecolumn, draftcls, 12pt]{IEEEtran}

\usepackage{url,comment,cite}
\usepackage{amsmath,amssymb,amsfonts,bbm,dsfont}
\usepackage{graphicx}
\usepackage{epsfig,epsf,psfrag,pstool,textcomp,tabularx}
\usepackage[usenames,dvipsnames]{xcolor}
\usepackage{subfigure}
\usepackage{float}
\usepackage{stfloats,mathtools}
\usepackage[algoruled,medskip,dontprintsemicolon,linesnumbered,Algorithm]{algorithm}
\usepackage[noend]{algpseudocode}
\usepackage{amsmath,mathtools}

\newtheorem{theorem}{Theorem}

\newcommand{\Cc}{\mathcal{C}}

\newcommand{\Lc}{\mathcal{L}}
\newcommand{\Pc}{\mathcal{P}}

\newcommand{\Wc}{\mathcal{W}}
\newcommand{\Tc}{\mathcal{T}}

\newcommand{\Sc}{\mathcal{S}}
\newcommand{\Zc}{\mathcal{Z}}

\newcommand{\ls}[1]
  {\dimen0=\fontdimen6\the\font
   \lineskip=#1\dimen0
   \advance\lineskip.5\fontdimen5\the\font
   \advance\lineskip-\dimen0
   \lineskiplimit=.9\lineskip
   \baselineskip=\lineskip
   \advance\baselineskip\dimen0
   \normallineskip\lineskip
   \normallineskiplimit\lineskiplimit
   \normalbaselineskip\baselineskip
   \ignorespaces
}

\begin{document}

\title{ Distributed Downlink Power Control for\\Dense Networks with Carrier Aggregation}
\vspace{-0mm}
\author{Zana Limani Fazliu,~\IEEEmembership{Student Member,~IEEE}, Carla-Fabiana Chiasserini,~\IEEEmembership{Senior Member,~IEEE}, Gian Michele Dell'Aera, Enver Hamiti

\thanks{Z. Limani Fazliu and C.-F. Chiasserini  are with Politecnico di Torino, Italy, email: name.lastname@polito.it.  C.-F. Chiasserini is also a Research Associate with CNR-IEIIT, Italy. G.M. Dell'Aera is with Telecom Italia, Italy, email: gianmichele.dellaera@telecomitalia.it. E. Hamiti is with the Faculty of Electrical and Computer Engineering, University of Prishtina, Kosovo, email: enver.hamiti@uni-pr.edu. }
\vspace*{-10mm}}

\maketitle

\maketitle
\thispagestyle{empty}\pagestyle{plain}

\vspace{-0mm}
\begin{abstract}
Given the proven benefits cell densification brings in terms of capacity and coverage, it is certain that 5G networks will be even more heterogeneous and dense. However, as smaller cells are introduced in the network, interference will inevitably become a serious problem as they are expected to share the same radio resources. Another central feature envisioned for future cellular networks is carrier aggregation (CA), which allows users to simultaneously use several component carriers of various widths and frequency bands. By exploiting the diversity of the different carriers, CA can also be used to effectively mitigate the interference in the network. In this paper, we leverage the above key features of next-generation cellular networks and formulate a downlink power setting problem for the different available carriers. Using game theory, we design a distributed algorithm that lets cells dynamically adjust different transmit powers for the different carriers. The proposed solution greatly improves network performance by reducing interference and power consumption, while ensuring coverage for as many users as possible. We compare our scheme to other interference mitigation techniques, in a realistic large-scale scenario. Numerical results show that our solution outperforms the existing schemes in terms of user throughput, energy and spectral efficiency.
\end{abstract}

\vspace{0mm}
\section{\label{sec:intro}Introduction} 

Densification of wireless cellular networks, by overlaying smaller cells over the traditional macrocell, is seen as an inevitable step in enabling future networks to support the expected increase in data rate demand.  As we move towards 5G, networks will become more heterogeneous as services will be offered via various types of points of access (PoAs). Indeed, besides the traditional macro base station, it is expected that users will be able to access the network through WiFi access points, small cell (i.e., micro, pico and femto) base stations, or even other users when device-to-device communications are supported. This approach will improve both the capacity and the coverage of current cellular networks, however, since the different PoAs are expected to fully share the available radio resources, inter-cell interference as well as the interference between the different tiers will pose a significant challenge \cite{andrews-5g}.

Future networks are also expected to support carrier aggregation (CA), which allows the simultaneous use of several component carriers (CCs), in order to guarantee higher data rates for end users. Downlink transmissions over the CCs will be characterized by different values of maximum output power depending on the type of PoA, and each carrier will have an independent power budget \cite{3gpp-trca}. Thus, since CCs may belong to different frequency bands, they may have also very different coverage areas and impact in terms of interference, due to both their different transmit power level and their propagation characteristics. 

Currently, three main approaches have been proposed to address the interference problem in dense networks: per-tier assignment of carriers, Enhanced Inter Cell Interference Coordination (eICIC), which has been adopted in LTE-A systems, and downlink power control. Per-tier assignment of carriers simply implies that in CA-enabled networks, each tier is assigned a different CC so as to nullify inter-tier interference \cite{lp-abs}. eICIC includes techniques such as Cell Range Expansion (CRE) to incentivize users to associate with micro base stations, and Almost Blank Subframes (ABS), i.e., subframes during which macrocells mute their transmissions to alleviate the interference caused to microcells.  Algorithms to optimize biasing coefficients and ABS patterns in LTE heterogeneous networks have been studied in, e.g., \cite{eicic-alg}, however they do not address CA.  Also, modifications to the eICIC techniques that allow macro base stations to transmit at reduced power during ABS subframes have been proposed in \cite{lp-abs}. In this paper we do not consider a solution within the framework of eICIC or its modifications, rather we use them as comparison benchmarks for the solutions we propose. 

We adopt instead the third approach, which consists in properly setting the downlink transmit power of the different CA-enabled PoAs so as to avoid interference between different tiers.  We propose to leverage the diversity in the component carrier coverage areas to mitigate inter-tier interference by varying their downlink transmit power. Thus, we enable a wide range of network configurations which reduce power consumption, provide high throughput and ensure a high level of coverage to network users. This type of configurations have also been envisioned by 3GPP \cite{3gpp-ca}, however, unlike the current specifications, we aim at reaching such solutions dynamically and in-response to real traffic demand.

As envisioned in LTE-A systems and unlike most of previous work, we consider that each CC at each PoA has an independent power budget, and that PoAs can choose the transmit power on each carrier from a discrete set of values.  Therefore, our goal is to adequately choose a power level from a range of choices to ensure optimal network performance.  It is easy to see that the complexity of the problem increases exponentially with the number of cells, CCs and the granularity of the power levels available to the PoAs. In addition, if one of the objectives is to maximize the network throughput, the problem becomes non linear since transmission data rates depend on the signal-to-interference-plus-noise ratio (SINR) experienced by the users. It follows that an optimal solution requiring a centralized approach would be both unfeasible and unrealistic, given the large number of cells in the network.

We therefore study the above problem through the lens of game theory, which is an excellent mathematical tool to obtain a multi-objective, distributed solution in a scenario with entities (PoAs) sharing the same pool of resources (available CCs). We model each group of PoAs in the coverage area of a macrocell as a team so that we can capture both (i) cooperation between the macrocell and the small cells with overlapping coverage areas, and (ii) the competitive interests of different macrocells. The framework we provide however allows for straightforward extension to teams that include several macrocells.  We prove that the game we model belongs to the class of {\em pseudo-potential} games, which are known to admit pure Nash Equilibria (NE) \cite{pa-potential}. This allows us to propose a distributed algorithm based on best-reply dynamics that enables the network to dynamically reach an NE representing the preferred solution in terms of throughput, user coverage and power consumption. As shown by simulation results, our scheme outperforms fixed transmit power strategies, even when advanced interference mitigation techniques such as eICIC are employed.

\section{\label{sec:rel-work}Related work} 
While many papers have appeared in the literature on uplink power control, fewer exist on downlink power setting.  
Among these,  
\cite{coalitions_overlap}  uses  coalitional games to investigate  power and resource allocation in heterogeneous networks where cooperation between players is allowed.  Downlink power allocation in cellular networks
is modeled  in  \cite{hierarchical-competition} as a Stackelberg game, with macro and femto base stations competing to maximize their individual capacities under power constraints. Resource allocation in heterogeneous networks is also addressed in \cite{diaz-geometric} where the authors propose two possible solutions, a heuristic approach using simulated annealing and geometric optimization, while taking into account both the geometry of the network and load fluctuations. Interference  in densely deployed femtocell networks is addressed in \cite{lin-powadj} through proper power adjustment and user scheduling. The authors propose a heuristic distributed algorithm that adjusts the coverage radius of the femtocells and then schedules the users in a fair manner. However,  the algorithm applies only to femtocells, thereby missing out on many possible solutions offering both better energy efficiency and network throughput. A backhaul-aware approach is taken by the authors in \cite{sapountzis-downlink} where they propose an optimal user association scheme to mitigate interference, which takes into account the base station load, the backhaul load as well as backhaul topology. 

An energy efficient approach is instead proposed in \cite{hetnet-eff}. There, base stations do not select transmit power levels as we do in our work, rather they can only choose between on and off states. 
Maximizing  energy efficiency is also the goal of \cite{yang-eeff}, which however is limited  to the study of resource allocation and downlink transmit power in a two-tier LTE single cell.  In \cite{udn-saad}, in order to improve the energy efficiency of ultra-dense networks, the authors frame the problem of joint power control and user scheduling as a mean-field game and  solve it using the drift plus penalty (DPP) approach in the framework of Lyapunov optimization. Mean-field games are also used in \cite{zahrani} where the interference problem (both inter-tier and inter-cell interference) is formulated as a two-nested problem: an overlay problem at the macrocell level and an underlay problem at the small-cel level. In the overlay problem, the macrocell selects the optimal action first, to provide minimum service, while  the underlay problem is then formulated as a non-cooperative game among the small cells. The mean-field theory is exploited to help decouple a complex large-scale optimization problem into a family of localized optimization problems. 

We remark that the above papers address heterogeneous dense networks but, unlike our work, they do not consider CA support, which will be a fundamental feature of future cellular  networks and significantly changes the problem settings. Also, \cite{yang-eeff, coalitions_overlap, hierarchical-competition}  formulate a resource allocation problem that aims at distributing the transmit power among the available resources under overall power constraints. In our work, instead, 
we do not formulate the problem as a downlink power allocation problem,  rather as a  power setting problem at carrier level, assuming {\em each carrier has an independent power budget}. Additionally, while most of the previous work \cite{hetnet-eff, discrete-eeff, yang-eeff, diaz-geometric,lin-powadj} focus on the heterogeneous network interference problem only, using game theory concepts we  jointly address  interference mitigation, power consumption  and user coverage by taking advantage of the diversity and flexibility provided by the availability of multiple component carriers.  Finally, we  propose a 
solution that enables the PoAs to dynamically change their power strategies based on user distribution, propagation conditions and traffic patterns. 

To our knowledge, the only existing work that investigates downlink power setting in cellular networks with CA support is  \cite{joint-ra-ca}.  There,  Yu et al. formulate an optimization problem that aims at maximizing  the system energy efficiency by optimizing power allocation and user association. However, interference issues, which are one of the main challenges we address, are largely ignored in \cite{joint-ra-ca} as the authors  consider a non-heterogeneous,  single cell scenario.

\section{System model and assumptions\label{sec:system}}
We consider a CA-enabled two-tier dense network composed of macro  and  microcells, each controlled by different types of PoAs. The network serves a large number of CA-enabled user equipments (UEs), which may move at low-speed (pedestrian) or high-speed (vehicles).

To make the problem tractable, we partition the entire network area into a set of identically-sized square-shaped tiles, or zones, denoted by $\Zc$. From the perspective of downlink power setting, the propagation conditions within a tile from a specific PoA represent averages of the conditions experienced by the UEs within the tile. Note that the tile size can be arbitrarily set, and represents a trade-off between complexity and realism. The choice, however, must be such that, the number of users falling within a tile should not be too high, and the assumption that they experience similar channel conditions should hold. We will assume that tiles (i.e., the UEs therein) are associated with the strongest received reference power, although the extension to other, dynamic association schemes as well as to the case where a tile is served by multiple PoAs can be easily  obtained. For simplicity, the user equipments (UEs) in the network area are all assumed to be CA enabled. 
Note, however, that the extension to  
a higher number of tiers as well as to the case where there is a mix of CA-enabled and non CA-enabled UEs is straightforward. 
All cells share the same radio resources. In particular, a comprehensive set of component carriers (CC), indicated by $\Cc$, is available simultaneously at all PoAs (PoAs having at their disposal a subset of CCs is a sub-case of this scenario). 
Each CC is defined by a central frequency  
 and a certain bandwidth.  The central frequency affects the carrier's coverage area, as the propagation conditions deteriorate greatly with increasing frequency.

The level of transmit power irradiated by each PoA on the available CCs can be updated periodically depending on the traffic and propagation conditions in the served  tiles, or it can be triggered by changes in UE distribution or traffic demand. The update time interval, however, is expected to be substantially longer than a resource block (RB)\footnote{A resource block (RB) is the smallest resource unit  that can be allocated to a UE in LTE. It is 180 kHz wide and 1~ms long.}  allocation period, e.g., order  of hundreds of subframes. 
 Indeed, since downlink power setting is based on 
averaged values of reported CSI's over the tile, 
it is not imperative for a power setting scheme to 
constantly have  accurate CSI 
for each user; additionally, it is not necessary for the update period to be
aligned with the coherence time of the channel. 
The PoAs can choose from a discrete set of available power levels, including 0 that corresponds to switching off the CC. The possible power values are expressed as fractions of the maximum transmit power, which may vary depending on the type of PoA, i.e., $\boldsymbol{P}=\{0.1, 0.2,...,1\}$. As noted before, each CC at each PoA has an independent power budget.

\section{Game theory framework\label{sec:game}}

\begin{figure}
\centering
	\includegraphics[width=0.7\columnwidth]{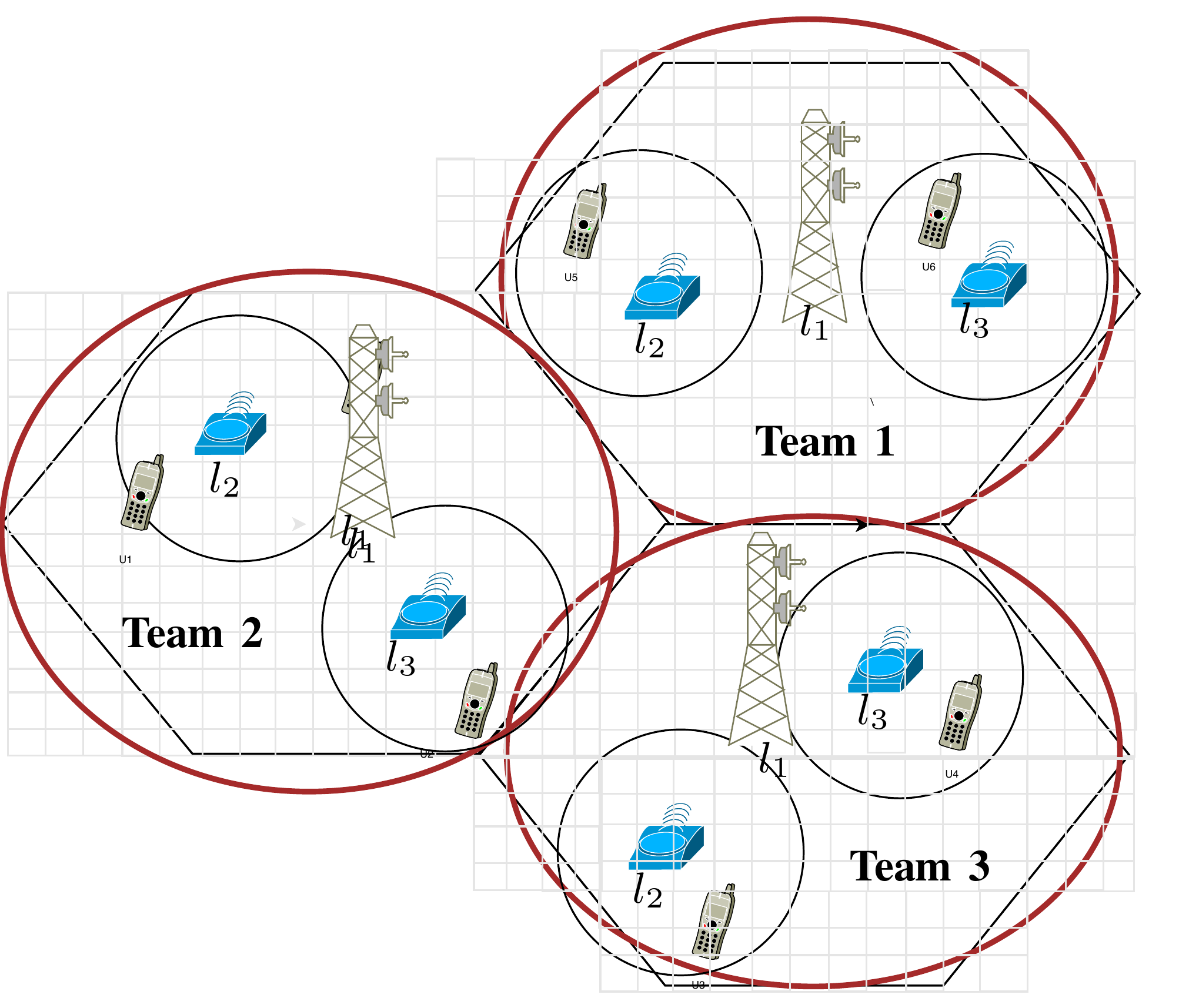}
\caption{\label{fig:net-model}Network model and teams. 
Team locations are denoted by $l_1, l_2, l_3$. Solid red lines represent team boundaries, while black solid lines represent coverage areas. Tiles are represented by grey squares.}
\vspace{-5mm}
\end{figure}
As mentioned before, game theory is an excellent tool to address complex problems, for which an optimal centralized solution might not be feasible.  In game theory, solutions to complex problems are usually reached by identifying the Nash Equilibria (NE) of the game; these are strategy profiles in which every player plays its best strategy, considering all other players' strategies fixed. Since none of the players have an incentive to unilaterally move from an NE strategy, such outcomes of the game are desirable, indeed they represent stable solutions which can be reached in a distributed manner. 
 Considering that the complexity of the carrier power setting problem increases exponentially with the number of PoAs, CCs and the granularity of the transmit power levels, we adopt a  game theoretic approach in order to derive low-complexity, distributed solutions  that lead to NEs and are  applicable in practice.

Specifically, we formulate the problem of power setting in  dense CA-enabled networks as a competitive game between {\em teams} of PoAs (see Fig.~\ref{fig:net-model}),  where each team wants to maximize its own payoff. Indeed, given the network architecture at hand,  PoAs within an overlapping geographical area have the common objective to provide the UEs under their coverage high data throughput. Thus, they  may choose to cooperate with each other in order to improve their individual payoffs as well as contribute to the ``social welfare'' of the team. Cooperation among such PoAs is beneficial especially since the inter-tier interference is most significant within the cell. 

It follows that teams will compete between each other for the same resources, each aiming at  maximizing their own benefits.  The game we model  and its analysis are detailed below. We note that the formulation can be easily extended to accommodate various team configurations and clusters of teams, each controlled by a central controller.

\subsection{Game model\label{subsec:game-definition}}

Let $G=\{\Tc,\Sc,\Wc\}$ be a competitive game between the set of players $\Tc$, where $\Sc$ is the comprehensive set of strategies available to the players  and $\Wc$ is the set of payoff functions. The objective of each player in the game is to choose a strategy such that it maximizes its payoff. The payoff function, in general, depends also on the strategies of the other players, thus a player must make decisions 
accounting for the strategies
the other players have selected. 
We now proceed to define the {\em players}, {\em strategies} and {\em payoff} functions in our scenario. 

\subsubsection{Players}
 As we mentioned in the previous section, we formulate the carrier transmit power setting as a competitive game between teams of PoAs. Hence the players in our competitive game are the {\em  teams}, each comprising a macro PoA and  the micro PoAs whose coverage areas geographically overlap with that of the macrocell. The terms {\em team} and {\em player} are used interchangeably throughout the paper. We denote the set of teams in our network as $\Tc=\{t_1,...,t_{T}\}$, where $T$ is the number of teams. We assume that the team members exchange information between each other, and that the macro PoA  plays the role of  team leader, i.e., it makes the decisions for all team members in a way that maximizes the overall team benefits. Furthermore, we will refer to the PoAs forming a team $t$ as the {\em locations} of the team,  $\Lc_t=\{l_1,l_2,...,l_L\}$ where, for simplicity of notation, the number of locations within a team is assumed to be constant and equal to $L$. Such a generalization is particularly useful since the interference caused within the team depends also on the relative position between the different PoAs. We indicate the set of tiles under the coverage area of a particular location $l$ by $\Zc_l$, and their union, denoting the comprehensive set of tiles of the team, by $\Zc_t$.   In addition, we use $E_l$, $E_z$ and  $E_t$ to denote the number of UEs under the coverage of location $l$, tile $z$, and team $t$, respectively, with $E_t=\sum_{l\in\Lc_t}E_l=\sum_{z\in\Zc_t}E_z$.

\subsubsection{Strategies}
Each team, comprising  a set of locations, 
has to decide which  transmit power level to use (out of the possible values in $\boldsymbol{P}$), at each one of those locations and for each of the available carriers $\Cc=\{c_1,c_2,...,c_C\}$. 
It follows that the strategy selected by a team $t$, $\boldsymbol{s^t}$, is an $L\times C$ matrix, where each $(l,c)$ entry indicates the power level chosen from set $\boldsymbol{P}$, 
to be used at location $l$ on carrier $c$. Consequently, the strategy set available to a team will be composed of all possible combinations of power levels, locations and carriers.

\subsubsection{Payoff functions} In game theory payoff functions are used to model the objectives of the players, usually expressed in terms of utility and cost, when choosing between different available strategies. Since network throughput is an important performance metric in cellular networks, it is natural that the utility of each team in our scenario is defined as a function of the data rates it can serve to its UEs. The data rate a UE obtains is closely linked to the SINR it experiences, which depends on the transmit power chosen by the serving location (PoA), the CC that is used and the transmit power levels chosen by neighboring locations. Assuming that all UEs within the same tile experience the same amount of interference, for each team we can first define an interference matrix of size $|\Zc_t| \times C$, denoted by $\boldsymbol{I^t}$.  Each entry in the matrix indicates the interference experienced by UEs in tile $z$ on carrier $c$, which is caused by other teams:  
\vspace{-2mm}
\begin{equation}
I^t_{z,c}(\boldsymbol{s^{-t}}) = \sum_{t'\in\Tc \wedge t'\neq t}\sum_{l'\in\Lc_{t'}}s^{t'}_{l',c}a_{l',z,c}
\label{eq:interference}
\end{equation} 
where $\boldsymbol{s^{-t}}$ represents the strategies adopted by all teams other than $t$, $s^{t'}_{l',c}$ is the power level (the strategy) of team $t'$ for location $l'$ on carrier $c$, and $a_{l',z,c}$ is the attenuation factor ($0\leq a_{l',z,c}\leq1$) related to the signal transmitted from location $l'$ on $c$ and received by the UEs in tile $z$. The attenuation values are pre-calculated using the urban propagation models specified in \cite{itu}.
The SINR  at tile $z$, when served by location $l$ in team $t$, is:
\vspace{-2mm}
\begin{equation}
\gamma_{z,c}^t=\frac{s^{t}_{l,c}a_{l,z,c}}{N+\sum_{l'\in\Lc_t \wedge l'\neq l}a_{l',z,c}s^t_{l',c}+I^t_{z,c}}
\label{eq:SINR}
\end{equation}
where $N$ represents the average noise power level. Note that, besides $N$ and $I^t_{z,c}$, we have an additional term at the denominator,  which stands for the intra-team interference and indicates the sum of all power received from the locations within the same team, other than  location $l$. 

Then the utility of each team can be defined as a function of the individual tiles' SINR values. 
In particular, the sigmoid-like function has been  often used for this purpose in uplink power control \cite{pa-sigmoid}. We note that this function is  suited to capture also the utility in  downlink power setting, as it has features that closely resemble the realistic relationship between the SINR and the data rate.  
We therefore adopt the sigmoid function proposed in \cite{pa-sigmoid}, as the utility function of each (tile, carrier) duplet  in the team, and write the team utility as:
\vspace{-2mm}
\begin{equation}
u^t(\boldsymbol{s^t},\boldsymbol{s^{-t}}) = \sum_{l\in\Lc_t} \sum_{z\in\Zc_l}\sum_{c\in\Cc}\frac{E_z}{E_t \left(1+e^{-\alpha(\gamma^t_{z,c}-\beta)} \right)} \,. \label{eq:team-utility-sigmoid}
\end{equation}
The sigmoid function in Eq.~(\ref{eq:team-utility-sigmoid}) has two tunable parameters, $\alpha$, which controls the steepness of the function, and $\beta$, which controls its centre.  They can be tweaked to 
best meet the scenario of interest. In particular, the higher the $\alpha$, the closer the function resembles a step function, i.e., the utility becomes more discontinuous with the increase of the SINR. The higher the $\beta$, the larger the  SINR for which a tile  obtains a positive utility. In our scenario, $\alpha$ and $\beta$ are set so that the resulting sigmoid-like function captures the relationship between SINR and throughput. 
In addition, 
the individual utility of each tile $z$ in team $t$ is weighted by the fraction of UEs
 covered by the team  in the tile ($E_z/E_t$) so as to give more weight to more populated tiles. This enables us to account for the user spatial distribution whenever this is not uniform over the network area.  

Next, we introduce a cost function to account for the interference and its detrimental effect, as well as for fairness in the service level to users.  We define a first cost component that aims at penalizing  teams who choose high power strategies, as:
$ \sum_{l\in\Lc_t}\sum_{c\in\Cc}\xi^t_{l,c}\bar{a}_{l,c}s^t_{l,c}$ 
where $\bar{a}_{l,c}$ is the link quality (i.e., attenuation) on carrier $c$ averaged over all  tiles served by location $l$, and  $\xi^t_{l,c}$ is the  price per received power unit for location $l$ and carrier $c$. This cost component increases with the increase in the chosen level of transmit power, however  it also accounts for the propagation conditions of the users served by the location. 
In other words, locations that have to serve UEs experiencing poor channel quality will incur a lower  cost, which ensures some level of fairness. The way the unit price, $\xi$, should be set is investigated in Sec. \ref{sec:price-set}.

The second term of the cost function further provides fairness in the network 
by  penalizing those strategies that leave  UEs without coverage. It is defined as
$\delta e_t$, 
where $\delta$ is a unit price paid for each unserved user and $e_t$ is the fraction of UEs within the team area that experience SINR levels below a certain threshold. We remark that since a macro PoA can communicate with the  micro PoAs in the macrocell, the team leader has knowledge of the UE density  under the coverage of its locations.  Thus, it can easily estimate the fraction of users, $e_t$, depending on the strategy  chosen for each of its locations ($\boldsymbol{s^t}$) as well as  on all other teams' strategies ($\boldsymbol{s^{-t}}$).  
The total  cost function is then given by: 
\vspace{-2mm}
\begin{align}
\pi^t(\boldsymbol{s^t},\boldsymbol{s^{-t}})= \sum_{l\in\Lc_t}\sum_{c\in\Cc}\xi^t_{l,c}\bar{a}_{l,c}s^t_{l,c}+\delta e_t \label{eq:fullcost}
\end{align}
where $\xi$ and $\delta$  indicate the weight that is assigned to each part of the cost function. 
Finally, we define the payoff of each team $t$ as the utility minus the cost paid:
\vspace{-2mm}
\begin{align}
w^t(\boldsymbol{s^t},\boldsymbol{s^{-t}})  = u^t(\boldsymbol{s^t},\boldsymbol{s^{-t}}) -\pi^t(\boldsymbol{s^t},\boldsymbol{s^{-t}})  \,.\label{eq:teampayoff}
\end{align} 

As mentioned, a team's goal is to maximize its payoff. Provided that the team is aware of the strategies selected by other teams, it can choose among its available strategies, the one that maximizes the payoff function.  We will refer to this strategy as   {\em{best reply}}. 
 
Moreover, to reduce both power consumption and the interference towards other teams, a team will select its best reply among strategies that maximize its payoff, as follows.\\
\noindent
{\em (i)}~Between strategies that are equivalent in terms of payoff, it will choose the one with the lowest total power, to reduce the overall power consumption. \\
\noindent
{\em (ii)}~When indifferent between  strategies with equal total power but assigned to different locations, it will select the strategy that  assigns higher power levels to micro PoAs that are closer to the centre of the cell, to minimize interference. \\
\noindent
{\em (iii)}~When indifferent with respect to the two above criteria,
it will choose the strategy that assigns higher power levels to higher frequency carriers, again, to minimize interference. 

\vspace{-3mm}

\subsection{Price setting}\label{sec:price-set}

The price parameter $\xi^t_{l,c}$ introduced in Eq.~(\ref{eq:fullcost}) is an important parameter which affects the nature of the game. To gain some insight into the possible values of $\xi$, we can start by considering a single carrier, single location scenario. 
We further simplify the scenario to consider one tile per location; dropping the superfluous notation, the team payoff becomes:
\vspace{-2mm}
\begin{align}
w^t
&= \frac{1}{\left(1+e^{-\alpha(\frac{as^t}{\mathcal{I}^t+N}-\beta)} \right)} - \xi^tas^t
\end{align}
where $\mathcal{I}^t(\boldsymbol{s^{-t}})$ indicates the interference determined by other teams' strategies. We also set $\delta=0$, since the two cost components  are independent of each other, therefore the second component bears no effect on the analysis of the first component. Differentiating with respect to the team's chosen strategy,  $s^t$, which now is scalar, and solving for $0$, we get:
\vspace{-0mm}
\begin{align}
e^{-2\alpha(\frac{as^t}{\mathcal{I}^t+N}-\beta)}-\left(\frac{\alpha }{\xi^t(\mathcal{I}^t+N)}-2\right)e^{-\alpha(\frac{as^t}{\mathcal{I}^t+N}-\beta)}+1 = 0 \,.
\end{align}
From the above expression, we can derive the strategy tha maximizes the payoff, which turns out to be a real and positive value only if the following condition is satisfied: 
\vspace{-2mm}
\begin{align}
\xi^t\leq \frac{\alpha}{4\left(\mathcal{I}^t+N\right)} \,.\label{eq-priceubound}
\end{align}
The last expression indicates that the price parameter $\xi^t$ is inversely proportional to the interference experienced by the UEs served by the team location. If the interference experienced by the UEs in the tiles served by the location increases, it is clear that the value of $\xi^t$ needs to be lowered in order to ensure that the chosen power is a positive value. 
This suggests that, in order to achieve high performing operational points for our network, a dynamic price setting is required, so that the teams can adapt to the changing interference, as other teams change their strategies.  Also, note that the interference experienced by users in a tile does
  not affect the interference experienced by users in other tiles;
  the same holds for  the interference experienced by users in the
  same tile, but using different carriers. This implies that the
  relationship between the price parameter and interference remains
  the same, even when more 
  carriers and more  
  tiles are considered. 

We further remark that aside from being dynamically updated depending on the value of the interference, the price must also be tailored individually for each team location. Indeed, the interference experienced by UEs served by a specific location $l$ depends not only on the strategies selected by other teams, but also on the topology of the network, i.e., the relative position and distance between the interfering locations and said UEs. A team leader can leverage the knowledge it has about its team topology to adjust the price parameter, according to each location's expected external interference coming from other teams, and the expected intra-team interference.

\begin{algorithm}
\begin{algorithmic}[1]
\Require $c$, $\boldsymbol{s_c}$, $t$ \label{prs-input}
\ForAll{$l\in\Lc_t$ }
\State $\bar{I}^{t}_{l,c}=0$
\ForAll{$z\in \Zc_l$}
\State Compute $I^t_{z,c}$ by using Eq.~(\ref{eq:interference}) \label{line:prs-extint}
\State $I^{int}_{z,c}=\sum_{l'\in\Lc_t\wedge l'\neq l}s^t_{l',c}a_{l',z,c}$\label{line:prs-tiint}
\State $\bar{I}^{t}_{l,c}=\bar{I}^{t}_{l,c}+\frac{E_z}{E_l}\left(I^t_{z,c}+I^{int}_{z,c}\right)$\label{line:prs-ovint}
\EndFor
\State $\xi^t_{l,c}=\frac{k\alpha}{\bar{I}^t_{l,c}}$\label{line:prs-price}
\EndFor
\end{algorithmic}
\caption{\label{alg:price-setting}Dynamic team price setting}
\end{algorithm}

How to dynamically update the price for each team location under general settings is shown in Alg.~\ref{alg:price-setting}.
The procedure takes into account both the external interference coming from the other competing teams, calculated in line \ref{line:prs-extint}, as well as the internal interference coming from the other locations of the team, calculated in line \ref{line:prs-tiint}. Once these values are obtained, the price parameter $\xi^t_{l,c}$ is updated in line \ref{line:prs-price} using $\xi^t_{l,c}=\frac{k\alpha}{\bar{I}^t_{l,c}}$, where $k$ is a weight factor used to indicate the importance we place on the cost function; higher $k$ values indicate that consuming less power will be given more consideration when selecting the best response. As a result, for higher $k$ we obtain overall lower best response values, and vice versa.  Note that $k\leq1/4$ must hold in order  to satisfy Eq.~(\ref{eq-priceubound}). Note that the initial price for each team location is determined given an initial strategy, $\boldsymbol{s_c}$, which can be any of the fixed strategies, and then updated every iteration or game. Although more frequent updates can be implemented, our results show that it is sufficient to update the price parameters once for each game run.

\subsection{Game analysis\label{subsec:game-analysis}}
To analyse the behaviour of the above-defined game, and discuss the existence of NEs, we rely on the definition of games of {\em strategic complements/substitutes with aggregation} as provided in~\cite{pa-potential,pa-strategic}. 

A game $\Gamma=\{\Pc,\Sc,\Wc\}$, where $\Pc$ is the set of players, and $\Sc$ and $\Wc$ are defined as above, is a game of {\bf{strategic substitutes}} with aggregation if for each player 
 $p\in \Pc$ there exists a best-reply function  $\theta_p:\boldsymbol{S^{-p}} \to \boldsymbol{S^p}$ such that:
\vspace{-2mm}
\begin{align}
1)&\quad \theta_p(I^p)\in \Theta(I^p)\label{eq:cond-1}\\ 
2)&\quad \theta_p\text{ is continuous in } \boldsymbol{S^{-p}}\label{eq:cond-2} \\
3)& \quad \theta_p(\hat{I}^p) \leq \theta_p(I^p), \quad \forall \hat{I}^p>I^p \,.\label{eq:cond-3}
\end{align}
$\Theta(I^p)$ is the set of best replies for player $p$  and $\boldsymbol{S^{-p}}$ is the Cartesian product of the strategy sets of all participating players other than $p$.  $I^p$ is an additive function of all other players' strategies, also referred to as the {\em aggregator} \cite{pa-strategic}: 
\vspace{-2mm}
\begin{equation}
I^p(\boldsymbol{s^{-p}}) =\sum_{p'\in\Pc, p' \neq p} b_{p'}s_{p'}\label{eq:aggregator}
\end{equation}
where $b_{p'}$ are scalar values.  
Condition 1) is fulfilled whenever the dependence of the payoff function on the other players' strategies can be completely encompassed by the aggregator. Condition 2), also known as the {\em continuity} condition, implies that for each possible value of $I^p$, the best reply function $\theta_p$ provides unique best replies.  Condition 3) implies that the best reply of the team decreases with the value of the aggregator. 

A game of {\bf{strategic complements}} with aggregation is identical, except for condition 3), which changes into:
\vspace{-2mm}
\begin{align}
\theta_p(\hat{I}^p) \leq \theta_p(I^p), \quad \forall \hat{I}^p<I^p \,,\label{eq:cond-4}
\end{align}
i.e., in the case of games of strategic complements, the best reply of the team increases with the value of the aggregator.

Next, we show the following important result. 
\begin{theorem}
{\em Our competitive team-based game $G$ 
is a game of {\bf strategic complements/substitutes with aggregation}. } 
\end{theorem}

\IEEEproof
See Appendix A.
\endIEEEproof

As a further remark to the above result, it is worth stressing that the cost introduced in Eq.~(\ref{eq:fullcost}) is an important function that determines whether the game is of strategic complements or substitutes. Indeed, if we consider the payoff to coincide with the utility function (i.e., $\xi=\delta= 0$),  a team's best reply will consist in increasing its transmit power as the interference grows, implying that the game is of strategic complements. This would lead to an NE in which all teams transmit at maximum power level, without consideration for the interference caused. 
Instead, imposing some $\xi>0$, the game will  turn into a game of strategic substitutes. This is because the first term of the cost function is linear with the received power, and hence increasing with the chosen strategies. Therefore, the payoff function will start decreasing once the increase in the chosen transmit powers does not justify the price the team has to pay. Note that, throughout the paper,  we will consider  $\xi>0$ , therefore our game is of strategic substitutes. 
Imposing some $\delta>0$ (i.e., activating the second cost component), the relationship between transmit power and cost becomes more complicated but it does not change the nature of the game: the fraction of unserved UEs within the team will be high 
for 
very low power strategies, then it will decrease as the transmit power is increased, and increase again as the strategies chosen cause high intra-team interference. In other words, the second cost component  strengthens the trend in the payoff function imposed by the utility for increasing interference in presence of  low power strategies. For those mid-level strategies that ensure good coverage, it does not affect the cost function.  Instead, it resembles the behavior of the first cost component for high power strategies, as it is still able to discriminate against high power strategies that may harm the system performance.  

Main results from~\cite{pa-potential,pa-strategic} and references therein show that games of strategic complements/substitutes with aggregation belong to the class 
of {\em{pseudo-potential games}}, which are known to admit pure Nash Equilibria.
Another important result that holds for such games with a discrete set of strategies is that, thanks to the continuity condition in Eq.~(\ref{eq:cond-2}),  convergence to an NE is ensured by best reply dynamics \cite{pa-strategic,pa-potential}. 

\section{The power setting algorithm\label{sec:algo}}

We now use the above model and results to build a distributed, low-complexity scheme 
that enables efficient downlink power setting on each CC.
We first consider 
a single carrier  and show that it 
converges to the best NE among the possible ones, in terms of payoff. We aim for an NE because it is the only solution of the game which the participating teams can reach independently, although it may not be the most optimal one in terms of utility. We then extend the algorithm to the multiple-carrier case and discuss its complexity.

\subsection{Single-carrier scenario\label{subsec:single-carrier}}

Let us first focus on a  single carrier and consider two possible borderline strategies that a team may adopt: the {\em max-power} strategy in which all locations transmit at the highest power level, and the {\em min-power} strategy in which all locations transmit at the lowest available power level greater than 0. As shown in our previous study \cite{us-wowmom}, it transpires that the {\em min-power} always outperforms the {\em max-power} in a multi-tier dense scenario.
We therefore devise a procedure that should be executed by each team leader (macro PoA), in order to update the locations' downlink power setting, either periodically or upon changes in the user traffic or propagation conditions. It is based on the intuition that if teams are to start from the lowest possible strategy (i.e., zero transmit power), then the overall transmit power and interference would increase incrementally as teams play their best replies sequentially. Thus the game would converge to the NE with the lowest overall transmit power, which, as we argue below, would be preferred in terms of social welfare. To do that, all teams initialize their transmit power to zero, and sequentially run the Best-reply Power Setting (BPS) algorithm reported in Alg.~\ref{alg:single-cc-br}.

We refer to the single execution of the BPS algorithm by any of the teams as an iteration. Note that the order in which teams play does not affect the convergence or the outcome of the game, since all teams  start from the zero-power strategy. At each iteration, the leader of the team that is playing 
determines the strategy (i.e., the power level to be used at each location in the team) that represents the best reply to the strategies selected so far by the other teams. The team leader will then notify it to the neighboring team leaders that can be affected by this choice. BPS  will be run by the teams till convergence is reached, 
which, as shown in \cite{us-wowmom}, occurs very swiftly. Also, we remark that the strategies identified over the different iterations are not actually implemented by the PoAs.  Only the  strategies representing the game outcome will be implemented by the PoAs, which will set their downlink power accordingly for the current time period. 

In order to detail how the BPS algorithm (Alg.~\ref{alg:single-cc-br}) works, let us consider the generic $i+1$-th iteration and denote the team  that is currently playing by $t$. The algorithm requires as input the carrier $c$ at disposal of the PoAs and the strategies selected so far by the other teams, $\boldsymbol{s^{-t}_c}(i)$. Additionally, it requires the cost components weights $\xi$ and $\delta$, the SINR threshold  $\gamma_{min}$, used to qualify unserved users, and  the utility function parameters $\alpha$ and $\beta$. 
This latter set of parameters are calculated offline and provided to the teams by the network operator. 
The algorithm  loops over all  possible strategies in the strategy set of team $t$, $\boldsymbol{S^t_c}$. For each possible strategy, $\boldsymbol{s}$, and each location $l$ within the team, it evaluates the interference experienced by the tiles within the location area  (line~\ref{line:scc-interference}). This value is used to calculate the SINR and the utility (lines~\ref{line:scc-sinr}-\ref{line:scc-util}), then the first cost component is updated (line~\ref{line:power-cost}).  
In line~\ref{line:quality-cost1}, it is verified whether UEs in tile $z$ achieve the minimum SINR value. 
If not, the cost component $e_t$ is amended to include the affected UEs.
The overall team utility for each potential strategy 
$\boldsymbol{s}$ is obtained by summing over the individual tile utilities weighted  by the fraction of UEs present in each tile. We recall that such weight factor ensures that the UE distribution affects the outcome of the game accordingly. Once the utility and cost  are obtained, the team payoff corresponding to strategy $\boldsymbol{s}$ is calculated (line~\ref{line:payoff}). After this is done for all possible strategies, the leader  chooses the strategy $\boldsymbol{s^t}(i+1)$ that maximizes the team payoff. Note that, according to our game model, $\arg\max^{\star}$ in line~\ref{line:max} denotes the following operation: it applies the $\arg\max$ function and, if more than one strategy is returned, the best strategy is selected by applying the list of preferences in Sec.~\ref{subsec:game-definition}.

\begin{algorithm}
\begin{algorithmic}[1]
\Require $c$, $\boldsymbol{s^{-t}_c}(i)$, $\xi,\delta,\alpha,\beta,\gamma_{min}$ \label{line:scc-input}
\ForAll{$\boldsymbol{s}\in\boldsymbol{S^t_c}$}\label{line:scc-str}
\State Set $u^t(\boldsymbol{s},\boldsymbol{s^{-t}_c}(i))$, $w^t(\boldsymbol{s},\boldsymbol{s^{-t}_c}(i))$, $\pi^t(\boldsymbol{s},\boldsymbol{s^{-t}_c}(i))$, $e_t$ \hspace{-1mm} to \hspace{-1mm} 0 
\ForAll{$l\in\Lc_t$ {\bf and} $z\in \Zc_l$}
\State Compute $I^t_{z,c}$ by using Eq.~(\ref{eq:interference}) \label{line:scc-interference}
\State Compute $\gamma^t_{z,c}$ by using Eq.~(\ref{eq:SINR}) \label{line:scc-sinr}
\State $u^t(\boldsymbol{s},\boldsymbol{s^{-t}_c}(i)) \hspace{-1mm}\gets u^t(\boldsymbol{s},\boldsymbol{s^{-t}_c}(i))+\hspace{-1mm}\frac{E_z}{E_t\left(1+e^{-\alpha(\gamma^t_{z,c}-\beta)}\right)}$ \label{line:scc-util}
\State $\pi^t(\boldsymbol{s},\boldsymbol{s^{-t}_c}(i))\gets \pi^t(\boldsymbol{s},\boldsymbol{s^{-t}_c}(i))+\xi \bar{a}_{l,c}s_{l,c}$ \label{line:power-cost} 
\If{$\gamma^t_{z,c}\leq \gamma_{min} $} \label{line:quality-cost1}
\State $e_t\gets e_t+\frac{E_z}{E_t}$ \label{line:quality-cost2}
\EndIf
\State $\pi^t(\boldsymbol{s},\boldsymbol{s^{-t}_c}(i))\gets \pi^t(\boldsymbol{s},\boldsymbol{s^{-t}_c}(i))+\delta e_t)$ \label{line:power-cost2}
\EndFor
\State $w^t(\boldsymbol{s},\boldsymbol{s^{-t}_c}(i))\gets u^t(\boldsymbol{s},\boldsymbol{s^{-t}_c}(i))-\pi^t(\boldsymbol{s},\boldsymbol{s^{-t}_c}(i))$  \label{line:payoff}
\EndFor \label{line:scc-strend}
\State $\boldsymbol{s^{t}_c}(i+1)\gets \arg\max^{\star}_{\boldsymbol{s}}w^t(\boldsymbol{s},\boldsymbol{s^{-t}_c}(i))$\label{line:max}
\end{algorithmic}
\caption{\label{alg:single-cc-br}BPS Algorithm run by team $t$ at iteration $i+1$}
\end{algorithm}

\begin{theorem}
{\em When the NE is not unique, then  the BPS algorithm reaches the NE that maximizes the social welfare, i.e.,  the sum of individual payoffs. }
\end{theorem}

\IEEEproof
See Appendix B.
\endIEEEproof

\subsection{Multi-carrier scenario}

We now extend the BPS algorithm to the multi-carrier case.  As mentioned before, the team leader  has to decide on the power level to be used at each available carrier, at each location within the team. Thus the team strategy is no longer a vector, but an $L\times C$ matrix, each entry $(l,c)$ indicating the power level to be used for carrier $c$ at location $l$. 
A straightforward extension of Alg.~\ref{alg:single-cc-br} would imply that lines \ref{line:scc-str}--\ref{line:scc-strend} are executed for each element in the new extended strategy set. However, the new strategy set, depending on the number of carriers, may become too large and therefore make the algorithm  impractical to use in realistic scenarios. 

Analyzing the utility expression obtained in Eq.~(\ref{eq:team-utility-sigmoid}), we can note that since the carriers are in different frequency bands and have separate power budgets (as foreseen in LTE-A), the utilities secured at each carrier are independent of each other. In other words, the utility a team will get at one of the carriers, is not affected by the strategy chosen at another carrier. The same holds for the first cost component in Eq.~(\ref{eq:fullcost}). However, the overall payoff value is dependent on the interaction between carriers due to the second cost component.
Indeed, in networks with CA support, a UE can be considered unserved only if the SINR it experiences is below the threshold in all  carriers. 
In order to obtain a practical and effective solution in the multi-carrier scenario, we take advantage of the partial independence between the carriers, and run Alg.~\ref{alg:single-cc-br}  independently for each carrier, keeping the size of the strategy set the same as in the single-carrier scenario. Then, to account for the dependence exhibited by the second cost component, we set  the order in which the per-carrier games are played, using the order of preferences listed in Sec.~\ref{subsec:game-definition}. Since the teams prefer to use high-frequency carriers over low-frequency ones, due to their smaller interference impact, it is logical that the game is played starting from the highest-frequency carrier. It follows that  low-frequency carriers will likely be used to ensure coverage to UEs not served otherwise. 

Importantly, our algorithm is still able to converge to an NE, since surely none of the teams  will deviate from the strategies they chose at each carrier. Also, since the game for the lowest frequency carrier is played last, the number of served UEs  cannot be further  improved without increasing the power level on the other carriers, which we already know is not a preferable move as it has not been selected earlier.  Thus, although it does not search throughout the entire solution space as for the single-carrier scenario, the  procedure is still able to converge to an NE that 
provides a close-to-optimum tradeoff among throughput, user coverage and power consumption.  The results presented in \cite{us-wowmom}, obtained for toy scenarios, confirm that our scheme provides performance as good as that achieved by an exhaustive search in the strategy space.

\subsection{Complexity and overhead}
The complexity of the algorithm depends largely on the size of the strategy sets that are available to the teams, $\boldsymbol{S^t}$, since each team has to find the strategy which maximizes its payoff value by searching throughout the entire set. The set size depends on the number of discrete power levels available to the PoAs ($|\boldsymbol{P}|$), the number of locations in the team ($L$) and the number of CCs available at each location ($C$). In the single-carrier scenario, we have $|\boldsymbol{S^t}|=|\boldsymbol{P}|^L$, while in the multi-carrier scenario the size exponentially grows to $|\boldsymbol{S^t}|=|\boldsymbol{P}|^{LC}$, which is reduced to $|\boldsymbol{S^t}|=C|\boldsymbol{P}|^L$ by our approach. 

 In order to determine the downlink power setting, PoAs  leverage the feedback  they receive from their associated users on the channel quality that they  experience with respect to all PoAs within reference signal range.
 
These reports which are supported by current standards \cite{3gpp-36.331} occur approximately every 5~s.  Each location is expected to send these values, once per BPS update period,  to the team leader which will in turn run the BPS algorithm. We assume that PoAs within the same macrocell are interconnected, or at least connected to the macro PoA,  via, e.g.,  optical fiber connections, which allows for swift communication between them.  Thus, the overhead of control information flowing between PoAs within a team and their team leader is very limited and can be considered negligible. This is a reasonable assumption since it is expected that the architecture foreseen for future networks will allow PoAs that are geographically close to share a common baseband~\cite{andrews-5g}. In addition team leaders, i.e., macro PoAs, also need to  exchange their respective BPS outcomes at each iteration. Recall that, at each iteration, BPS produces the selected power level at each location and each carrier, which indicates that the team leaders need to exchange $L\times C$ integer values. In order to avoid additional overhead, team leaders can stop the broadcast of their BPS outcome, as soon as it is unchanged from the previous iteration.

\begin{figure}
\centering
\includegraphics[width=0.33\textwidth]{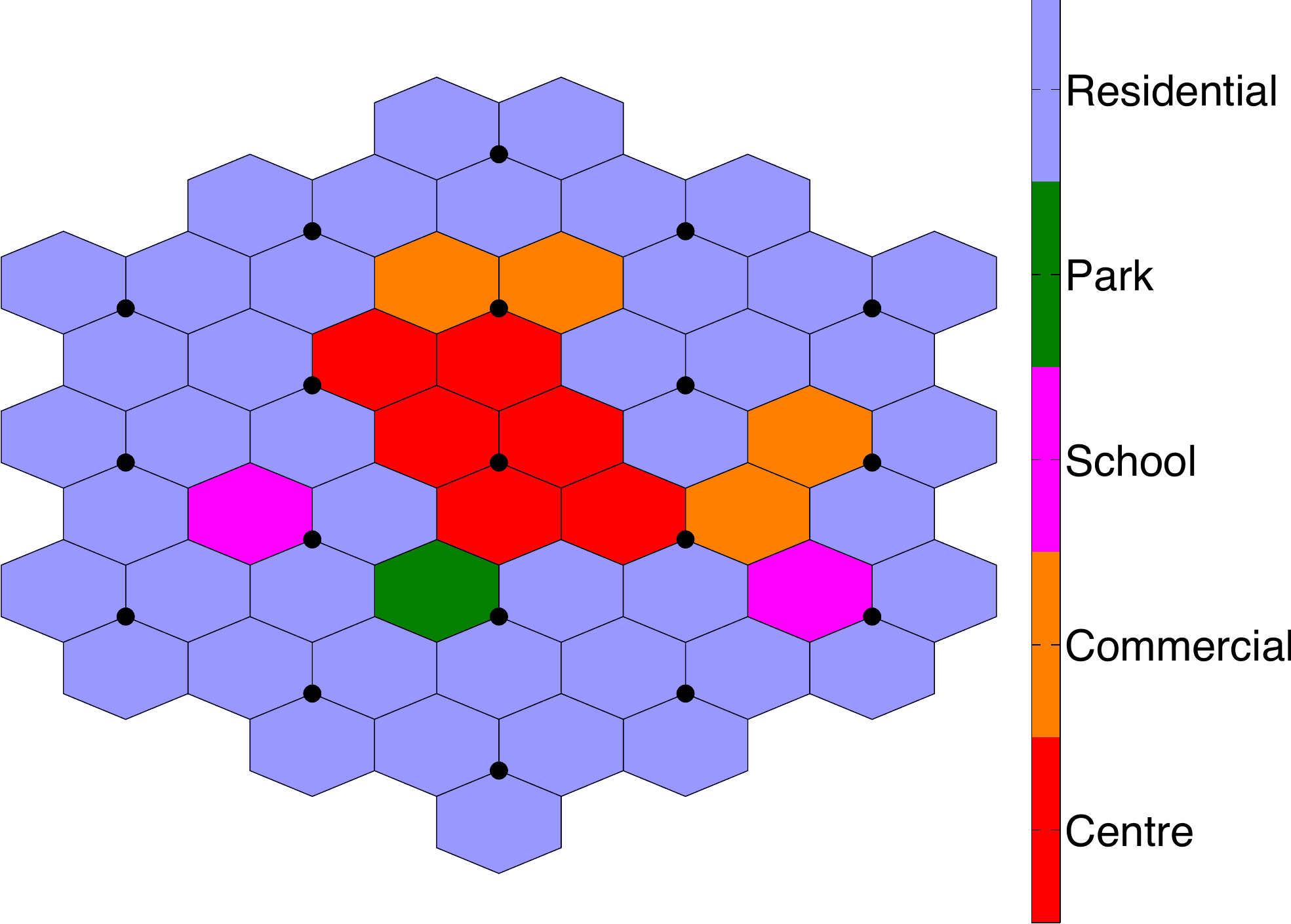}
\vspace{-3mm}
\caption{\label{fig:newsce}The network scenario and the different types of urban areas.}
\vspace{-3mm}
\end{figure}
\begin{figure*}
\centering
\includegraphics[width=0.25\textwidth]{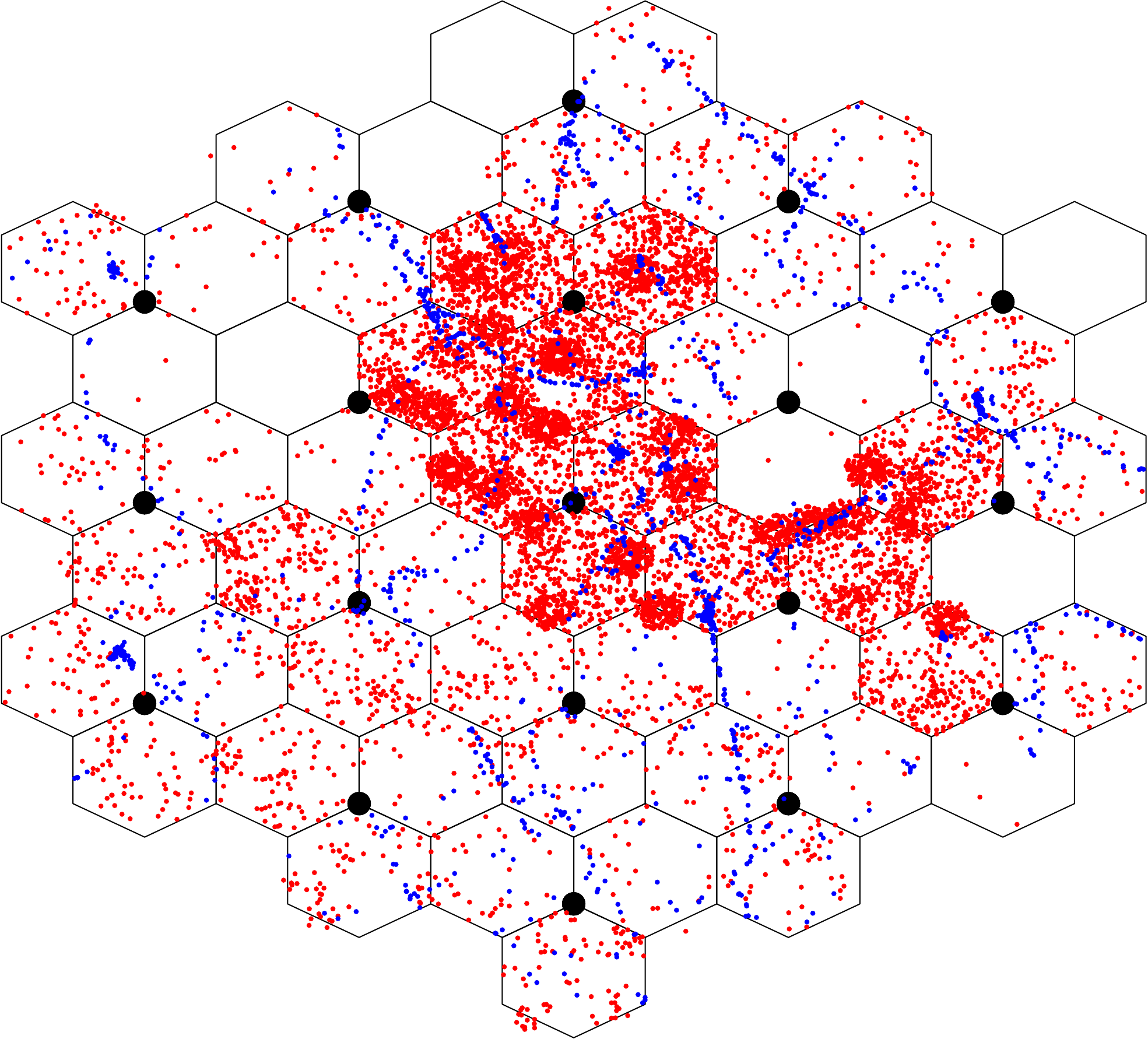}
\hspace{5mm}
\includegraphics[width=0.25\textwidth]{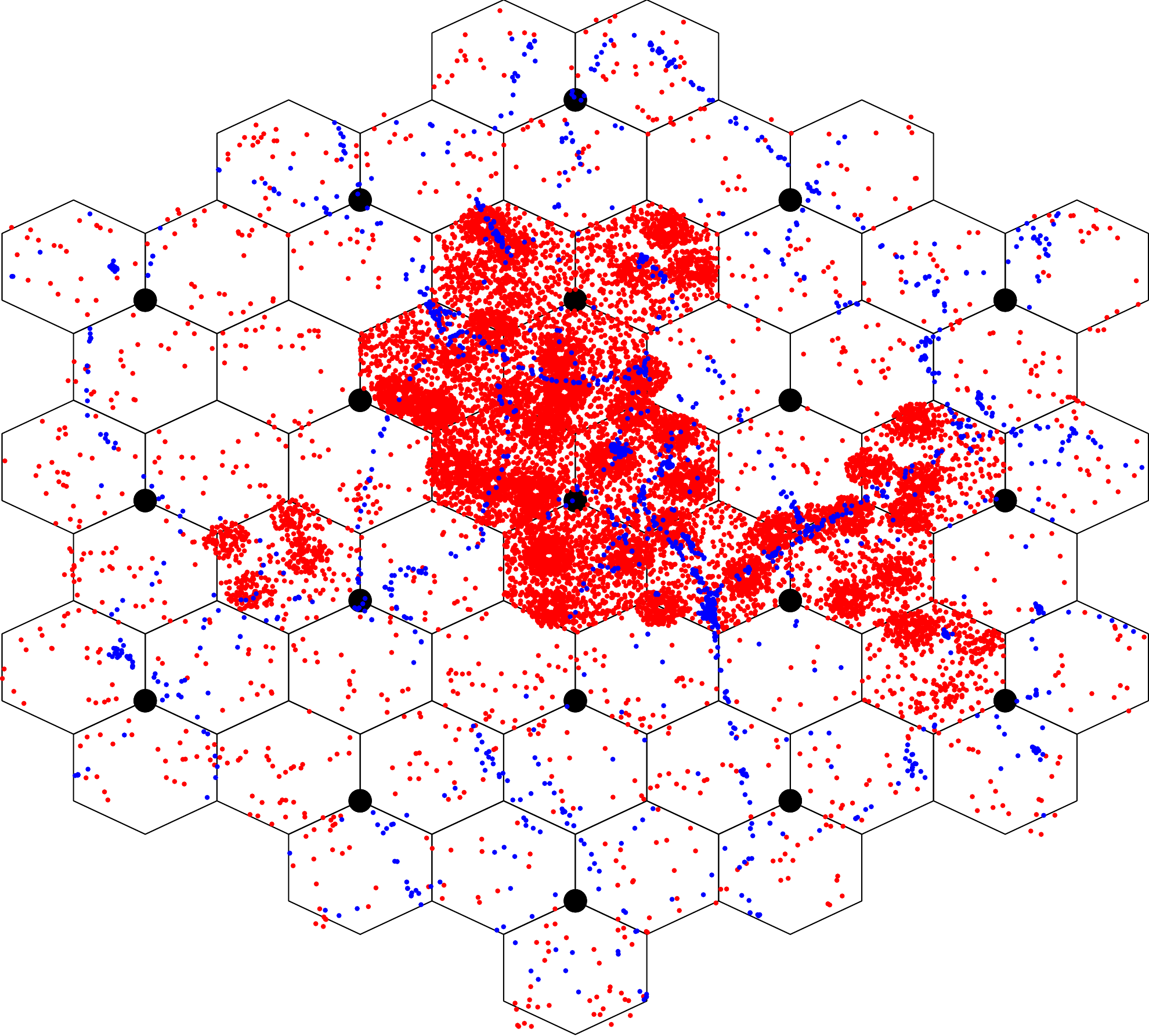}
\hspace{5mm}
\includegraphics[width=0.25\textwidth]{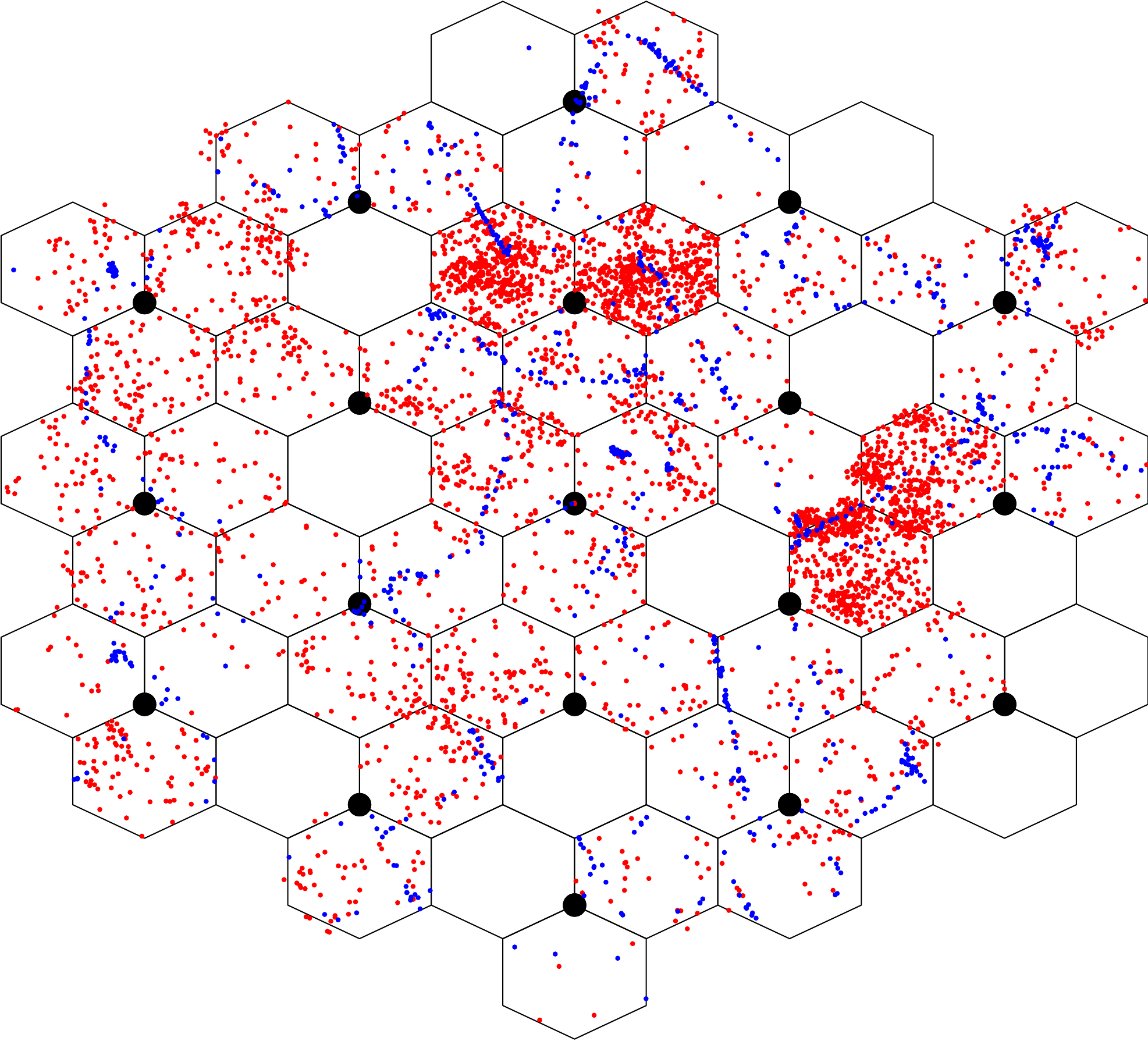}
\vspace{-3mm}
\caption{\label{fig:usdist}Snapshots of user distribution. The red dots represent pedestrian UEs, while the blue dots represent vehicle UEs. Left: Morning; Middle: Afternoon; Right: Evening.}
\vspace{-3mm}
\end{figure*}

\section{Performance evaluation\label{sec:peva}}

\begin{figure*}
\centering
\includegraphics[width=0.25\textwidth]{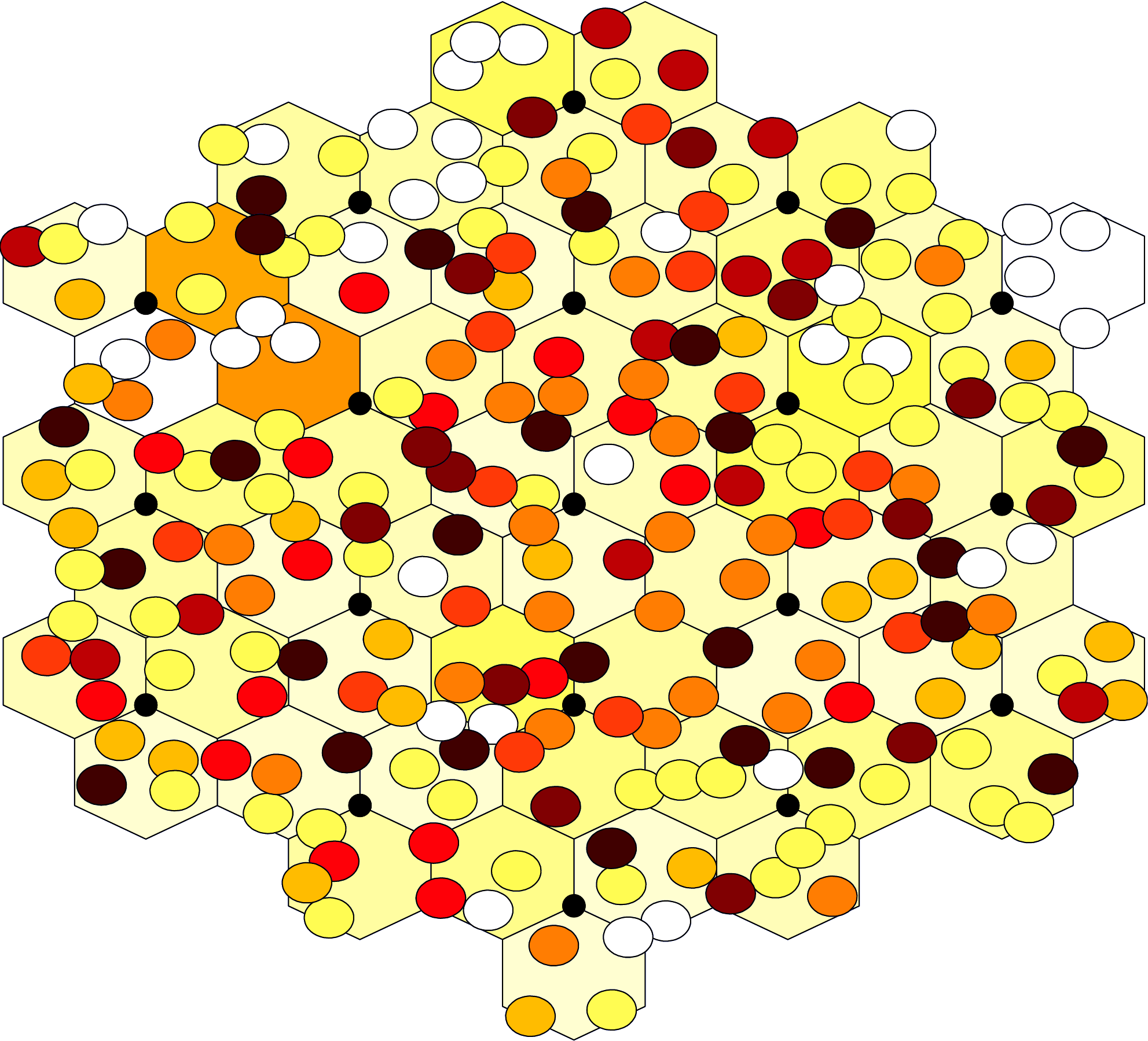}
\hspace{5mm}
\includegraphics[width=0.25\textwidth]{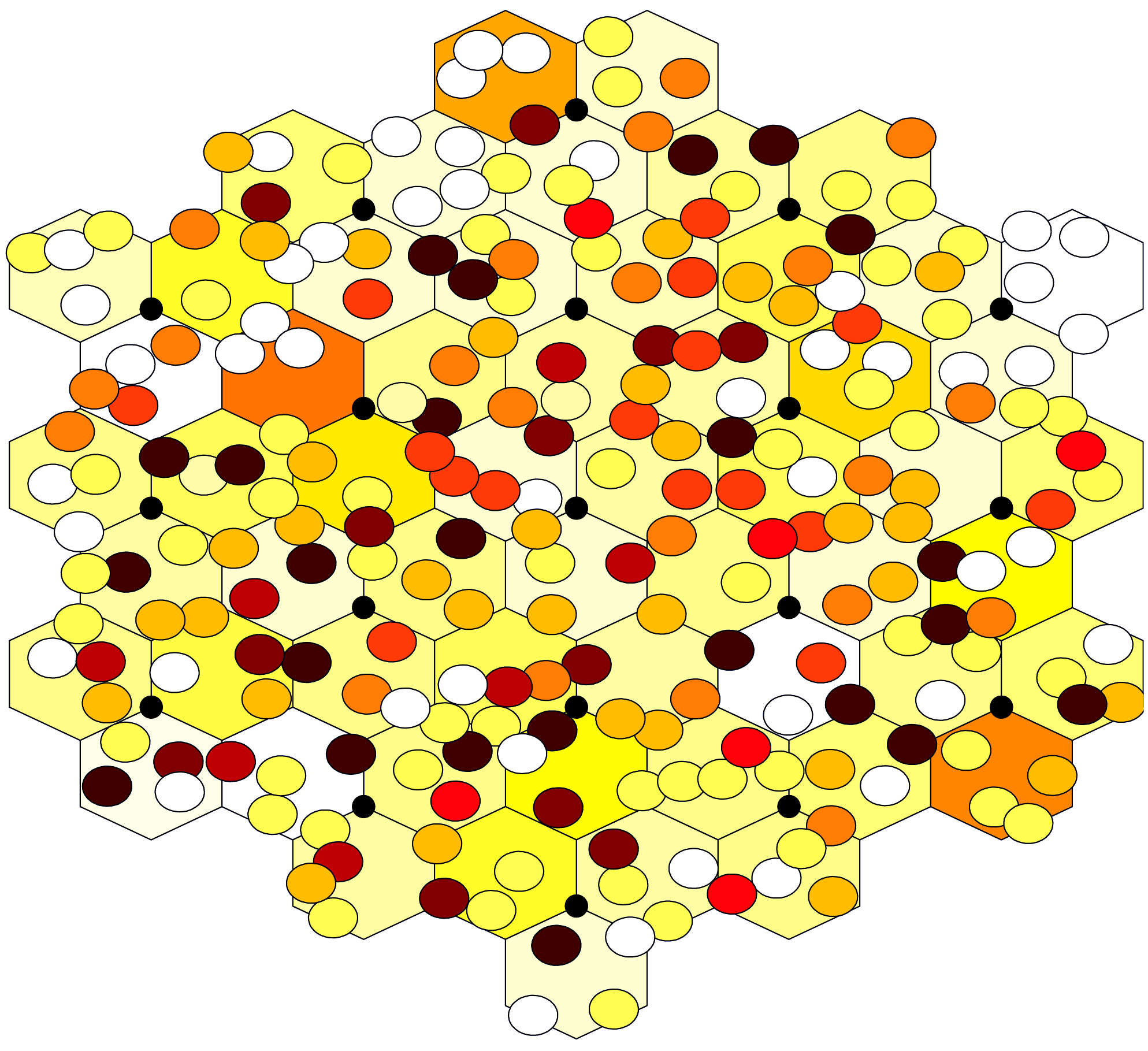}	
\hspace{5mm}
\includegraphics[width=0.25\textwidth]{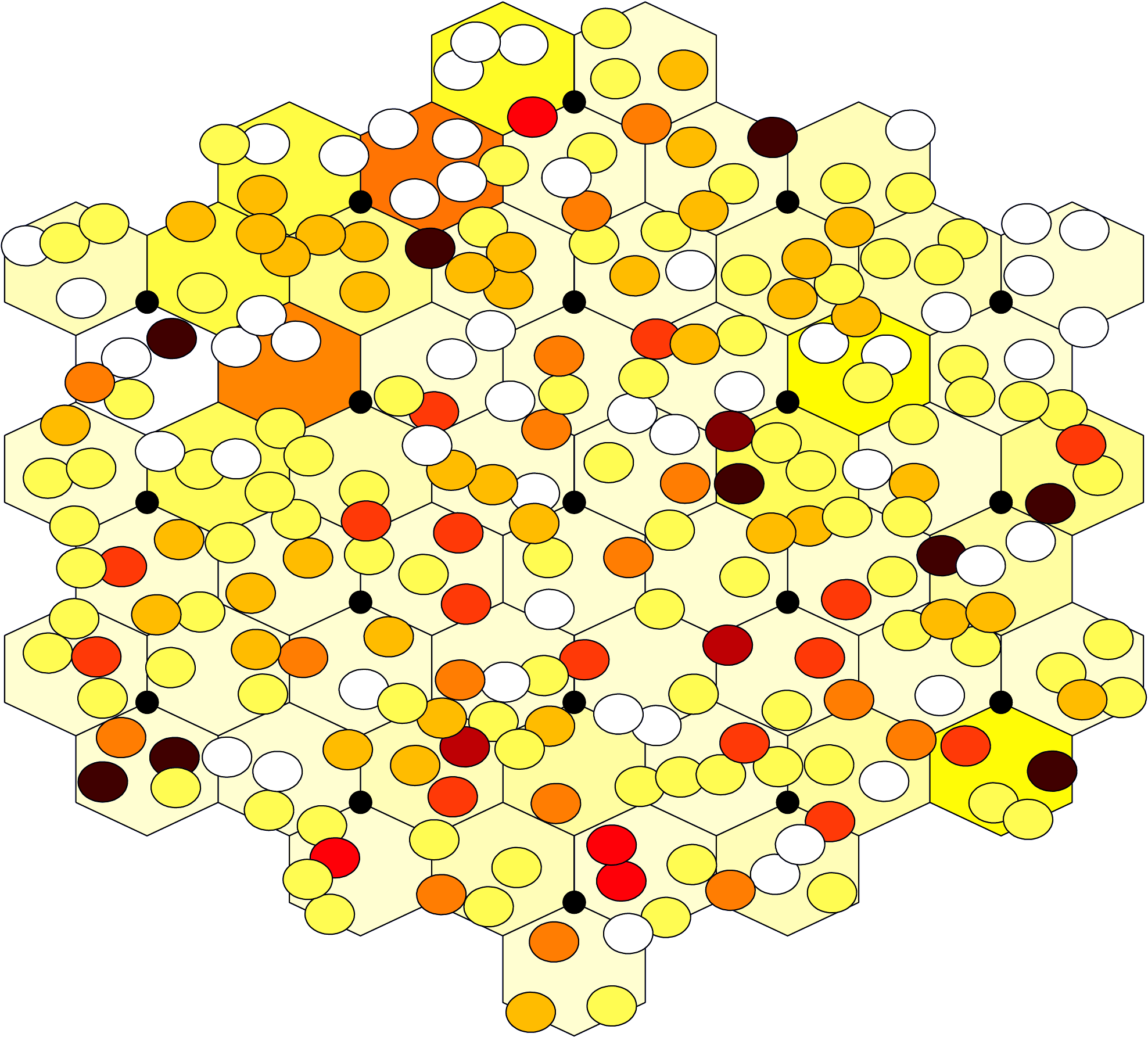}	
\vspace{-3mm}
\caption{\label{fig:necc}BPS achieved power strategy for the morning scenario.  Left: CC1; Middle: CC2; Right: CC3. }\vspace{-3mm}
\end{figure*}

We consider the realistic two-tier network scenario that is used within 3GPP for evaluating LTE networks~\cite{scenario}. The network is composed of 57 macrocells and 228 microcells.  Macrocells are controlled by 19 three-sector macro PoAs,  while  micro PoAs are deployed randomly over the coverage area so that there are 4 non-overlapping microcells per macrocell. The inter-site distance is set to 500~m. The overall network area is divided into 4,560 square tiles of equal size. The tile size was set so that an average of 2.5 and a maximum of 10 users fall within any tile, while ensuring that users  within a tile experience similar channel conditions. The PoAs are grouped into 57 five-location teams, each consisting of 1 macro PoA and 4 micro PoAs within its macrocell, unless stated otherwise.   Specifically, to make the scenario more realistic and comparable to an actual urban scenario, we divide the network coverage area into  five types of urban  areas:  city centre, residential area, commercial area, parks and school area, as shown in Fig.~\ref{fig:newsce}.  The UEs are also randomly dropped with varying density depending on the population density of the area type as well as time of the day (morning, afternoon or evening). Reference values for UE density were obtained using official population statistics of the city of Rome (Italy)~\cite{rome}, and then scaled to represent realistic values for cellular users of a single network provider. The UE densities were further scaled for the different urban areas and times of the day, using weights extracted from the data provided in the MIT Senseable City Lab project~\cite{mitlab}.  Note that, in addition, user density around micro PoAs is four times higher than over the rest of the macrocell. The mobility of pedestrian UEs was modeled using the random walk model, while the mobility of vehicular UEs was modelled using real mobility traces collected from taxi cabs in Rome \cite{trace-taxi}, assuming an average velocity of $30$~km/h. Snapshots of user distribution at different times of the day are shown in Fig.~\ref{fig:usdist}. The data traffic is simulated by generating download requests, whereby a random user requests to download a file which can be either video (file size: 1 Mb) or a generic file (file size: 500 kb), with equal probability. The number of requests per cell follows a Poisson distribution with a certain arrival rate $\lambda$ per cell, which varies depending on the urban area and time of the day. The final values obtained for user densities  and $\lambda$ are shown in Table. \ref{table:ue_dense}

\begin{table*}[t]
\caption{UE densities and cell request arrival rates\vspace{-2mm}}
\begin{center}
\begin{tabular}{ |c|c|c|c|c|c|}
\hline
 &City centre & Commercial area & School & Park & Residential area \\ \hline
 Baseline UE density [UE/msq] & 0.0245 & 0.0147 & 0.0074 & 0.0009 & 0.0009 \\ \hline
 Percentage of vehicles & 30\%& 5\%&5\%& 5\%&50\% \\ \hline
\multicolumn{6}{|l|}{\bf Density weights} \\ \hline
Morning (7-9 AM)&0.5&0.6&0.6&0.8&0.8\\ \hline
Afternoon (3-5 PM)&1&0.95&0.95&0.7&1\\ \hline
Evening (10 PM-12 AM)&0.08&0.5&0.01&0.5&0.6\\ \hline
\multicolumn{6}{|l|}{\bf Cell request arrival rates $\lambda$} \\ \hline
Morning (7-9 AM)&0.75&0.54&0.27&0.04&0.04\\ \hline
Afternoon (3-5 PM)&1.5&0.9&0.4&0.03&0.05\\ \hline
Evening (10 PM-12 AM)&0.12&0.45&0.005&0.02&0.03\\ \hline
\end{tabular}
\end{center}
\label{table:ue_dense}
\vspace{-4mm}
\end{table*}%

All UEs are assumed to be CA enabled. PoAs can use three CCs, each $10$ MHz wide, with the following central frequencies: 2.6 GHz (CC1), 1.8 GHz (CC2) and 800 MHz (CC3).  
We apply the ITU Urban Macro (UMa) model to calculate channel coefficients between macro PoAs and UEs, and the ITU Urban Micro (UMi) model for the channel between micro PoAs and users~\cite{itu}. In addition to path loss, we also consider shadowing effects, and, in the case of vehicular users, fast fading caused by the mobility.   SINR values are mapped on throughput using the look-up table in \cite{sinr-map}. The maximum transmit powers for macro  and micro PoAs are set at $20$~W and $1$~W, respectively~\cite{itu}. 
The game is played by all teams using the BPS algorithm for the multiple-carrier scenario. The power consumed by the network nodes is calculated using the power consumption model provided in \cite{earth}.
The sigmoid function parameters are $\alpha=1$ and $\beta=1$, which were selected as the most appropriate to model the relationship between the selected strategy and final user rate. 
The SINR threshold is set at $\gamma_{min}=-10$~dB, based on 
 \cite{sinr-map}. 
The value of the cost parameter $\xi$,  is calculated before running the BPS, using the dynamic pricing algorithm 
 in Sec.~\ref{sec:price-set}, with $k=0.25$. The power setting update period is set at 100~ms, which is considered sufficient from a practical perspective. Shorter update frequencies, as low as 10~ms, can also be implemented, provided that the delay incurred by the communication between macro PoAs is reasonable. 
However, while such short update time might make the algorithm more responsive to channel dynamics, we
 consider that longer update periods, such as 100 ms, perform excellently as confirmed by our results, while incurring significantly less signaling overhead.  Unless otherwise specified, the weight factor for the second cost component is set at
$\delta=0.6$.   Note that the values for $k$ and $\delta$  were chosen based on a numerical sensitivity analysis provided  in our previous study  \cite{us-wowmom}. 

The performance of the two algorithms is compared to the fixed strategy in which all PoAs transmit at highest power coupled with the eICIC technique, denoted as {\em eICIC} in the results. This combination was shown to perform best in our previous work \cite{us-wowmom} and is widely used in the literature and in practice.  eICIC is applied with CRE for microcells set at $8$~dB and 
macro PoAs downlink transmissions muted in  25\% of subframes (ABS). These values were chosen to represent the mid-range of those applied in the surveyed literature \cite{eicic-alg}. The underlying resource allocation is performed using the Proportional Fair (PF) algorithm. 

First, we take a look at the power setting strategies that the BPS algorithm produces. In Fig.~\ref{fig:necc}, we depict the averaged strategies reached through the BPS algorithm during the simulation period for the morning scenario.  The strategies chosen by the teams for each CC are differentiated using different shades, from white ({\em zero} power) to black ({\em maximum} power). Recall that the maximum power varies depending on the type of PoA. Hexagons represent the macro PoAs, while circles represent the micro PoAs.  The figure shows that CC1, i.e., the high frequency carrier, allows for higher transmit power levels to be used by both macro and micro PoAs, due to its low interference impact.  CC2 and, especially, CC3 are used to complement each other to ensure overall coverage. In general we see that low transmit power levels are preferred for macro PoA across all CCs, while for micro PoAs the chosen transmit power levels tend to be higher for higher frequency CCs such as CC1. It can also be noted that in highly concentrated areas such as the city centre and commercial areas, the micro PoAs tend to transmit at higher power levels, while macro PoAs at lower power levels. Such a strategy enables micro PoAs in these areas, which support most of the traffic demand, to transmit using a higher modulation coding scheme (MCS), which in turn implies higher bit rate and, hence,   throughput. In residential areas instead, traffic demand is lower and more spatially spread; thus, it is the macro PoAs that serve most of the traffic demand and therefore need to use higher power.

\begin{figure*}
\centering
\includegraphics[width=0.3\textwidth]{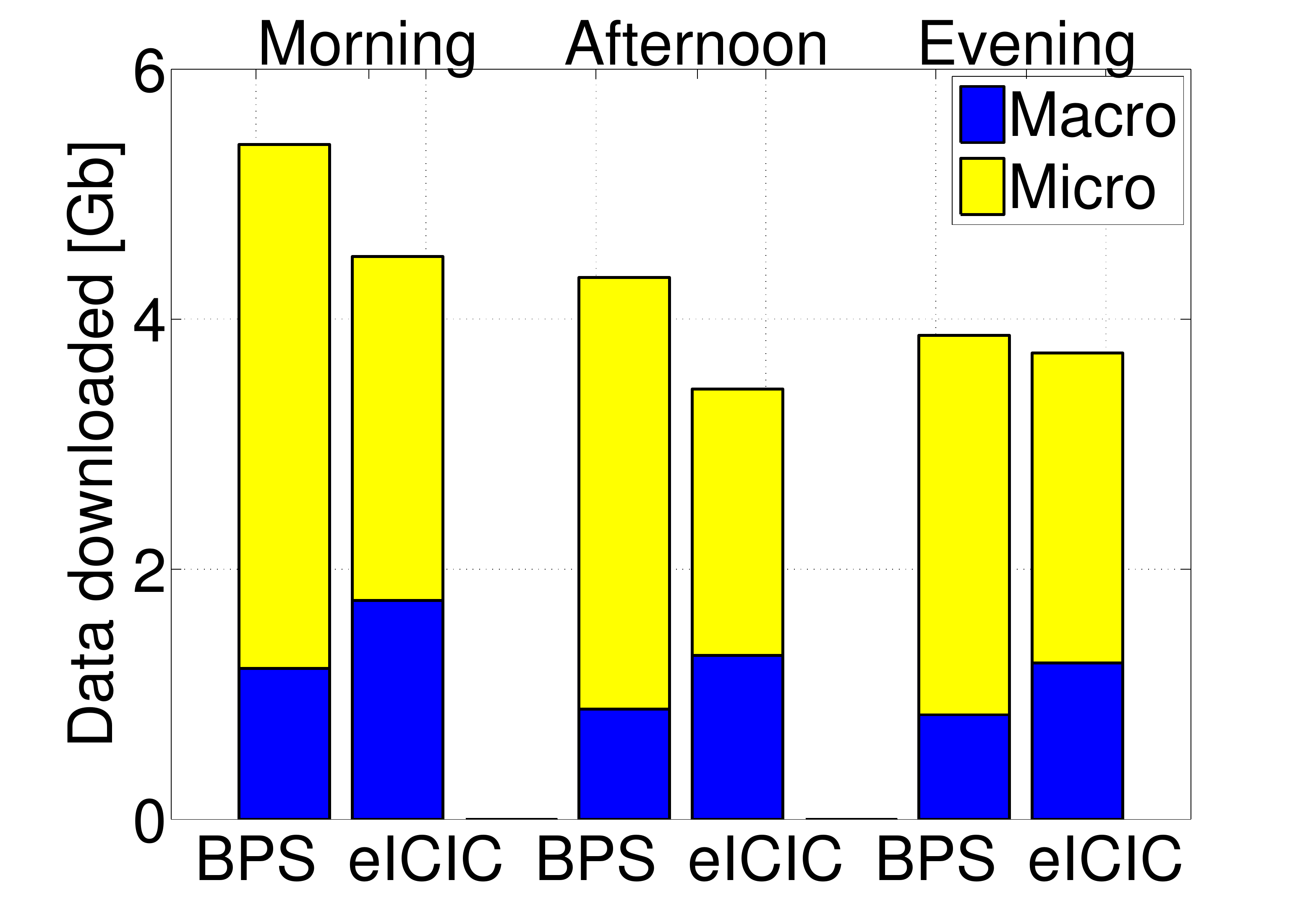}
\hspace{2mm}
\includegraphics[width=0.3\textwidth]{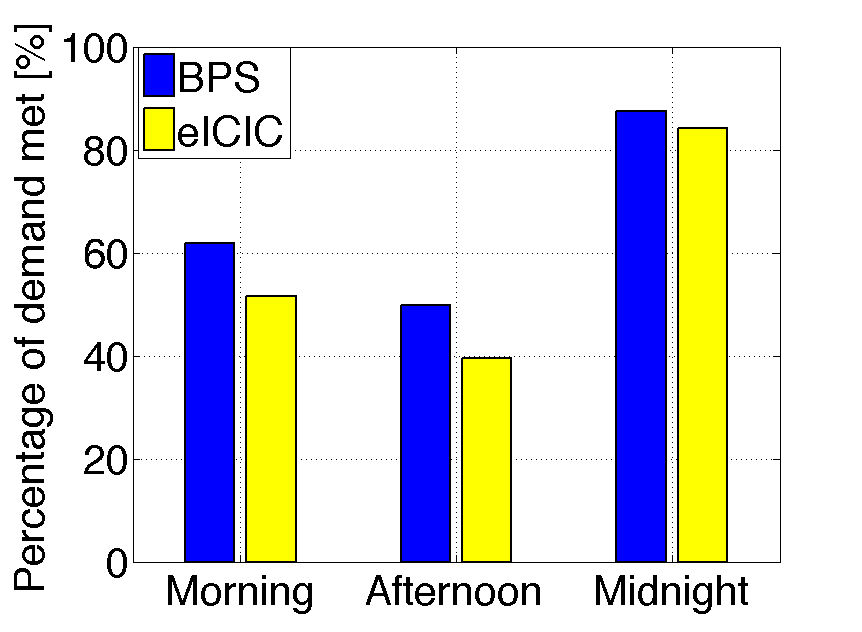}
\hspace{2mm}
\includegraphics[width=0.3\textwidth]{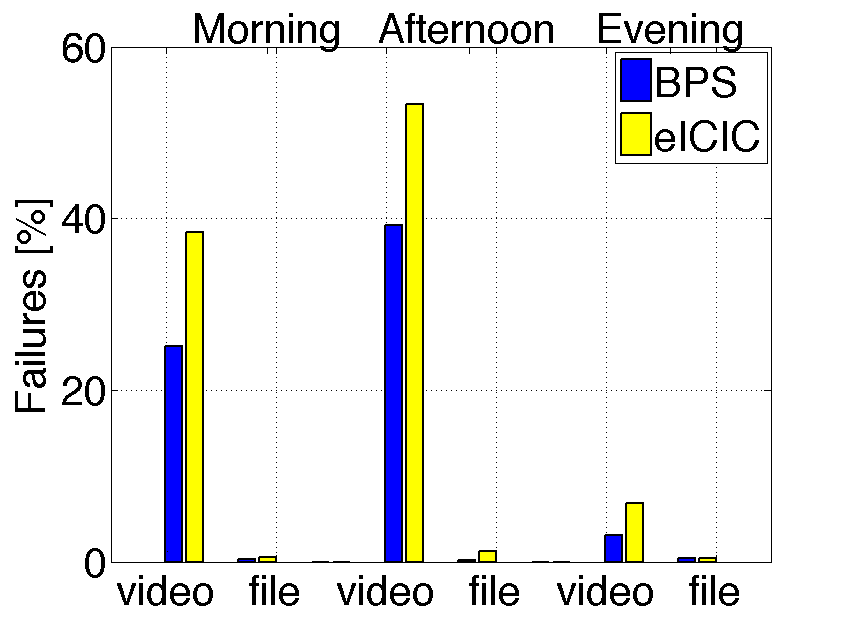}
\vspace{-3mm}
\caption{\label{fig:tot_content}Left: Total amount of downloaded content. Middle: Demand met. Right: Failed downloads per content type.} \vspace{-5mm}
\end{figure*}

In the following plots we show how the dynamically obtained power strategies outperform eICIC in some of the main performance metrics. Fig.~\ref{fig:tot_content} (left) shows that when BPS is employed the amount of data downloaded over the network is always higher, especially during high intensity periods like morning and afternoon.  BPS also improves the service experienced by the UEs in terms of demand met and percentage of failed downloads, as shown in Fig.~\ref{fig:tot_content} (middle) and (right). Note that for each type of file we have set a specific deadline (0.5 seconds for videos and 1 second for generic files), within which we expect the download to complete, otherwise it is considered a failed download. In Fig.~\ref{fig:tot_content} (middle), we show that during intensive periods, BPS improves the percentage of demand met across the entire network by around 10\%, and reduces the rate of failed downloads by approximately 15\%. It is clear that, as the traffic load intensifies in certain areas, which is the case in the afternoon scenario, both approaches have difficulties in managing the demand, however BPS does ensure an improvement, especially for video content, without applying any intelligent content-aware resource allocation techniques.  

\begin{figure*}
\centering
\includegraphics[width=0.3\textwidth]{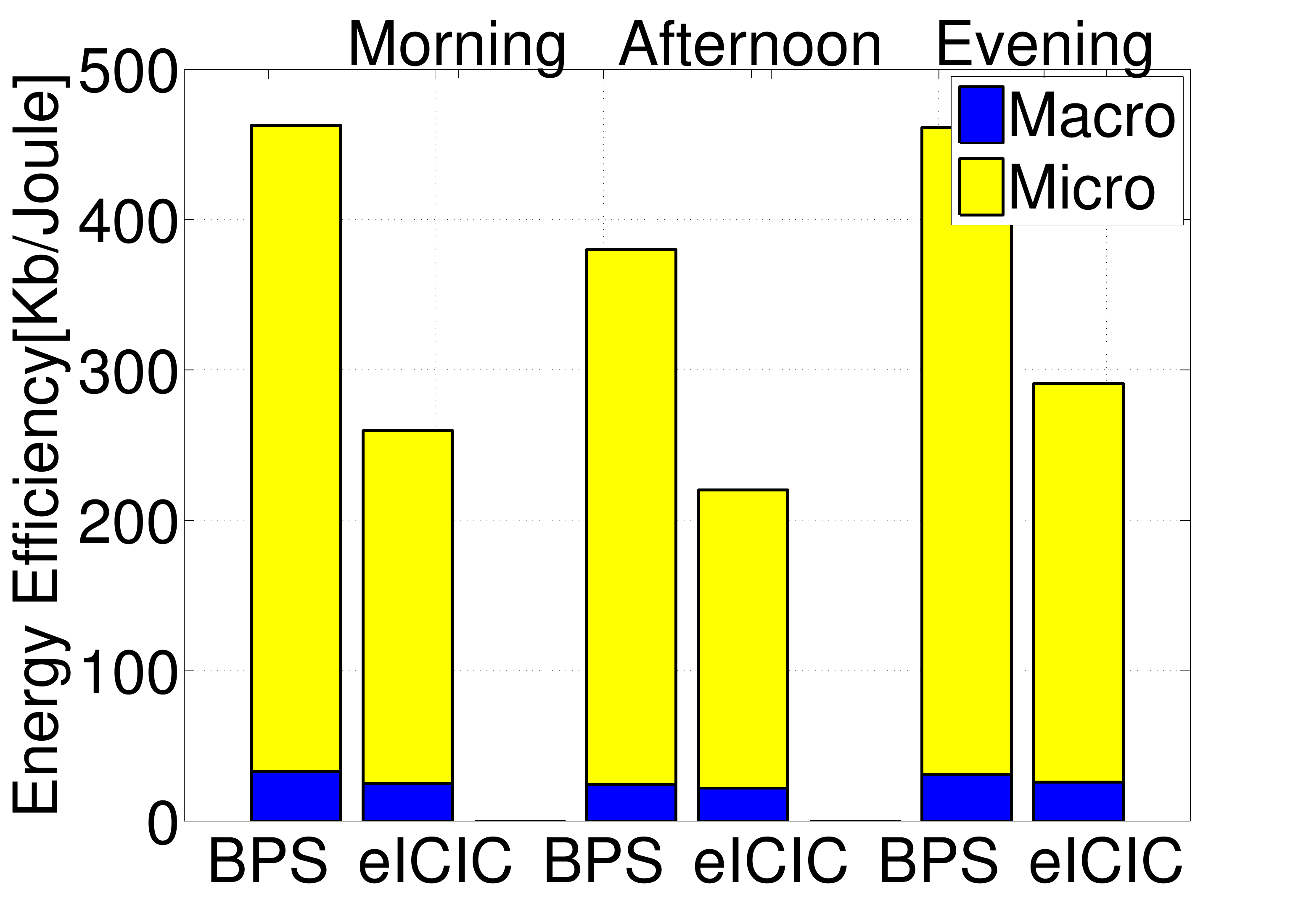}	
\hspace{2mm}
\includegraphics[width=0.3\textwidth]{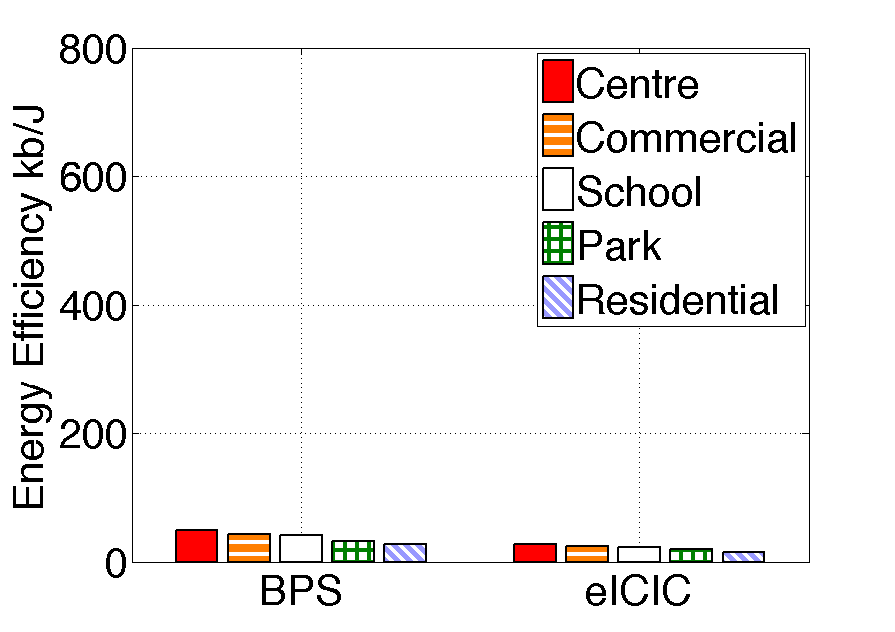}
\hspace{2mm}
\includegraphics[width=0.3\textwidth]{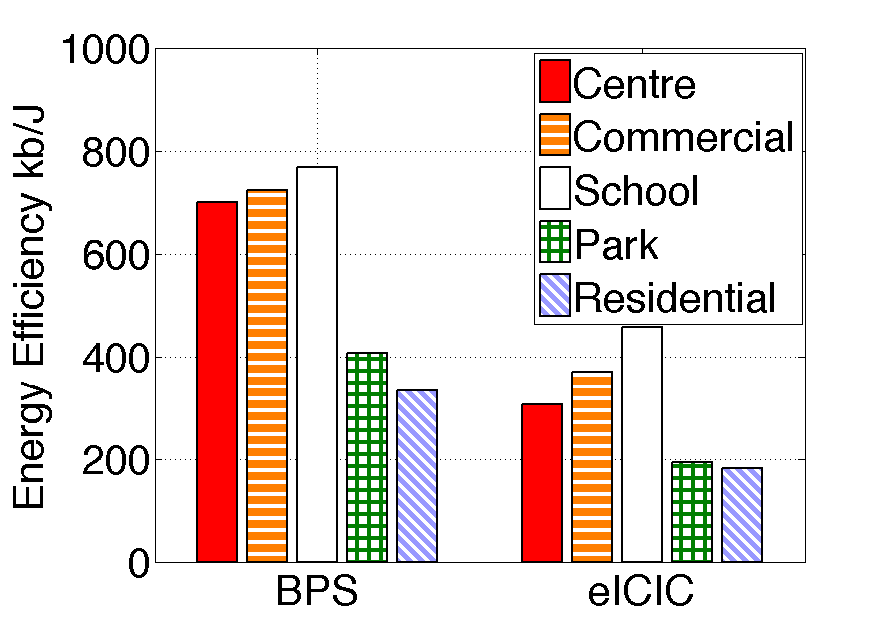}	
\vspace{-3mm}
\caption{\label{fig:eneff} Energy efficiency  in bits transmitted per joule consumed, at different times of the day (Left), and for different areas in the morning scenario, for  macro PoAs (Middle) and  micro PoAs (Right).}\vspace{-5mm}
\end{figure*}

While the difference is smaller in the evening when the traffic load decreases significantly, BPS still retains a considerable edge in energy efficiency (see Fig.~\ref{fig:eneff}). This is because BPS is able to serve higher amounts of data, while consuming significantly less power. From Fig.~\ref{fig:eneff} (left) it is clear that BPS improves the energy efficiency for both macro PoAs and micro PoAs, however the effect is more significant for the latter: the gain in energy efficiency for macro PoAs varies between 15 and 20\%, while for micro PoAs it can be as high as 100\% during the morning and it drops to around 60\% in the evening. BPS tends to choose lower transmit powers for macro PoAs, especially for dense areas with heavy traffic load, which significantly reduces the interference experienced by micro PoAs who are responsible for serving the  bulk of the data.  Indeed, if we look at the energy efficiency values for the  different areas, shown in Fig.~\ref{fig:eneff} (middle) and (right) for the morning scenario, it is clear that energy efficiency is highest in the city centre, commercial and school areas where the traffic load is more intense in the morning. Again, this is true for both macro and micro PoAs, but it is more significant for the latter.  

\begin{figure*}
\centering
\includegraphics[width=0.3\textwidth]{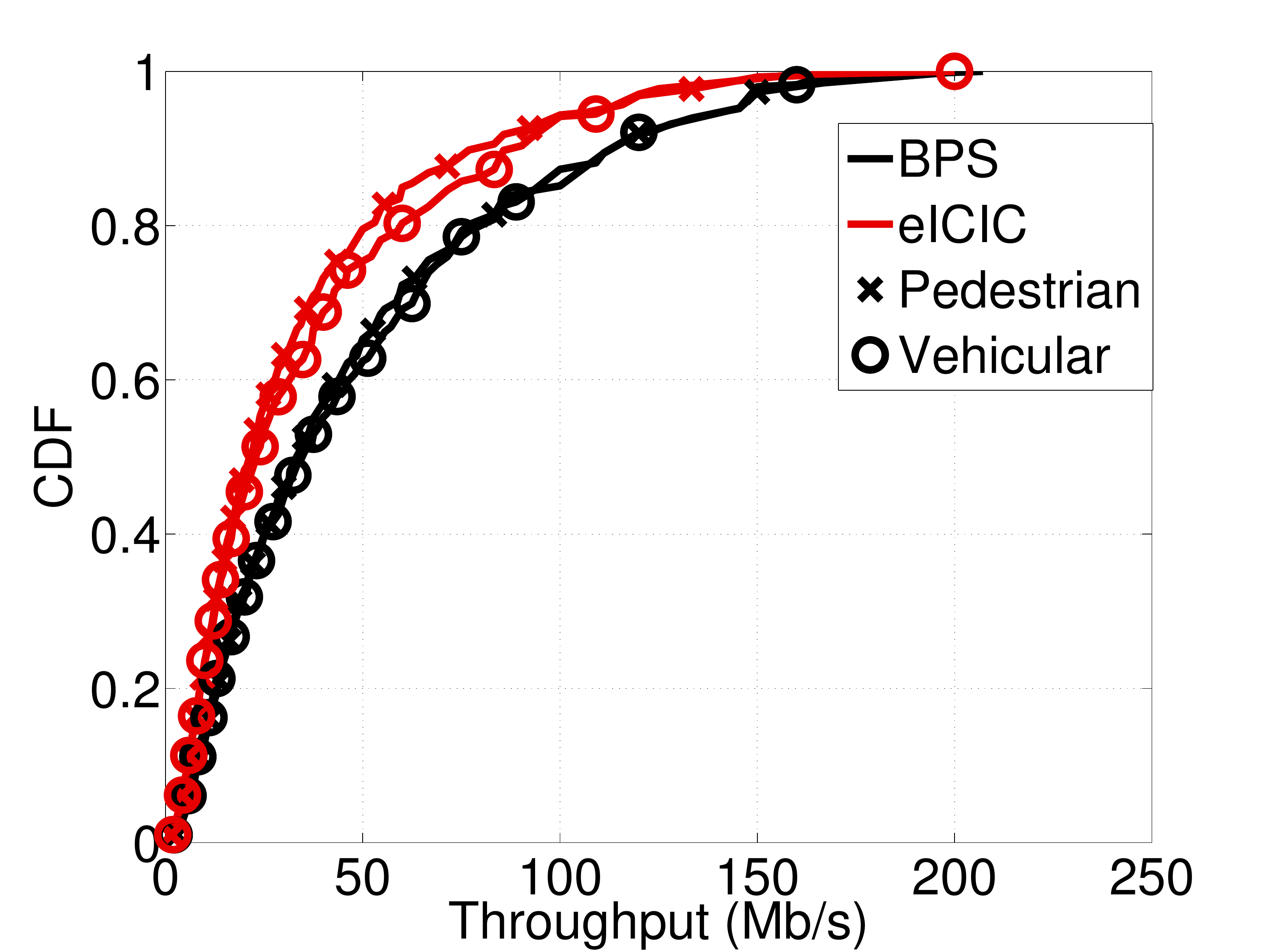}
\includegraphics[width=0.3\textwidth]{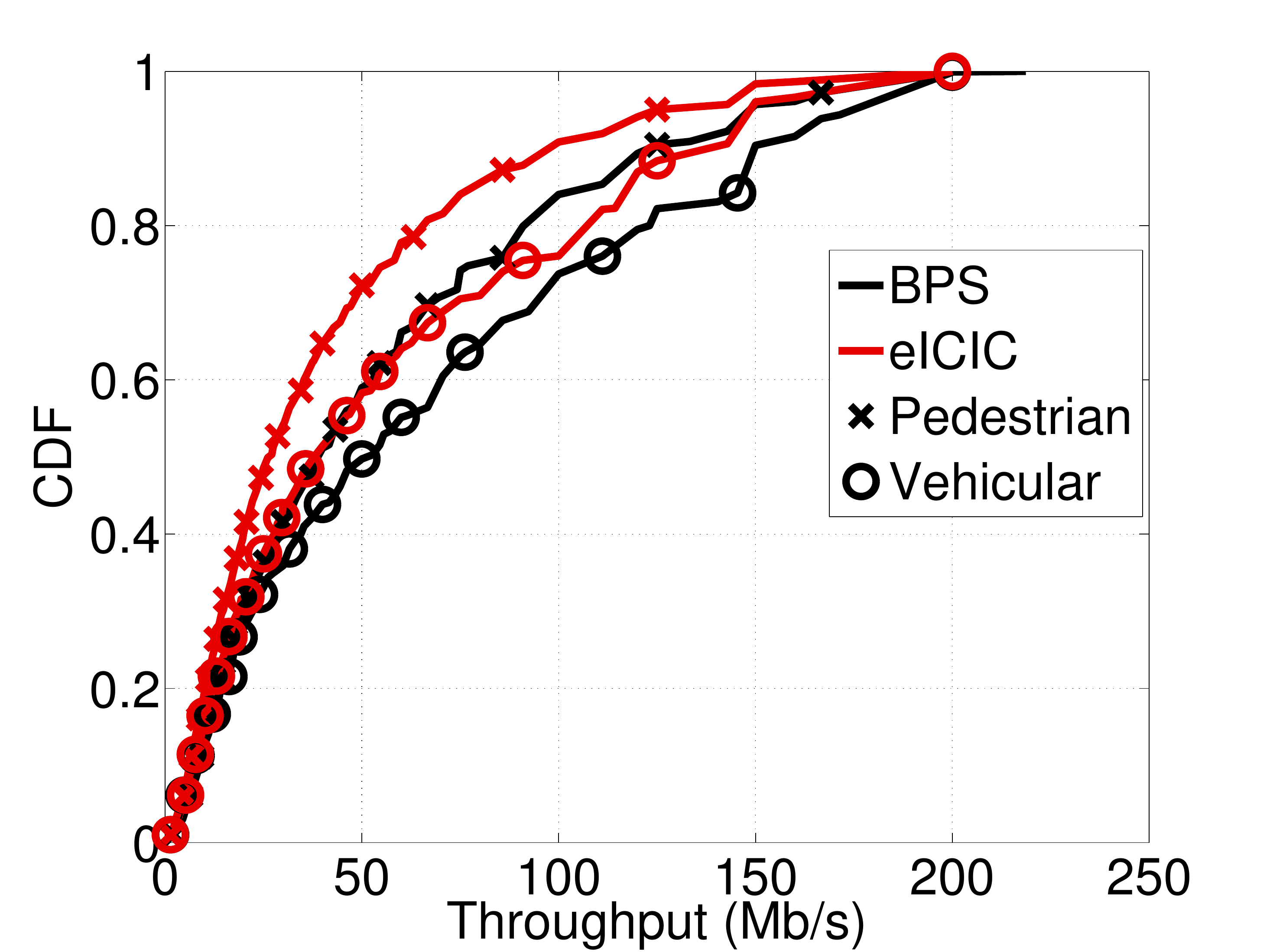}	
\includegraphics[width=0.3\textwidth]{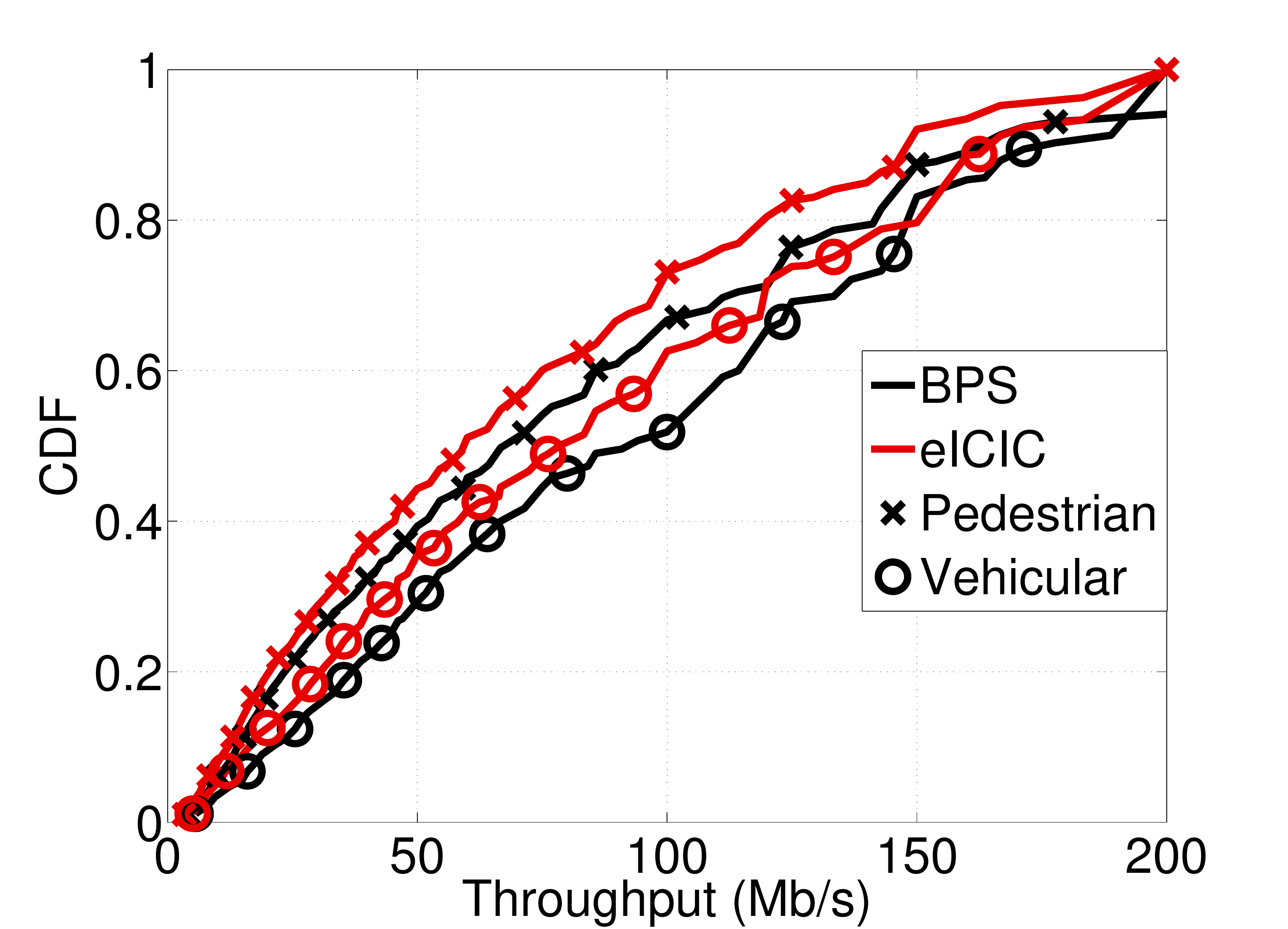}	
\vspace{-3mm}
\caption{\label{fig:cdf_veh_ped}CDF of the average user throughput achieved by pedestrian and vehicular UEs. Left: Morning; Middle: Afternoon; Right: Evening. }\vspace{-5mm}
\end{figure*}

In Fig.~\ref{fig:cdf_veh_ped} we look at the cumulative distribution function (CDF) of the achieved average user throughput at the different times of the day, differentiated for vehicular (circle) and pedestrian (cross) UEs. Note that, in general, BPS offers higher average user throughput for both types of traffic,  but the improvement compared to eICIC is more significant during peak hours. This is true especially for pedestrian UEs, who are concentrated in the high density areas with heavy traffic load. While it may look counterintuitive, vehicular UEs tend to have better average throughput. The reason is that most of the vehicle UEs tend to be spread in the residential area where the traffic demand is lower, and they tend to be situated in well covered areas. These two factors influence the performance more than the fast-fading effects.

\begin{figure*}
\centering
\hspace{-5mm}
\includegraphics[width=0.33\textwidth]{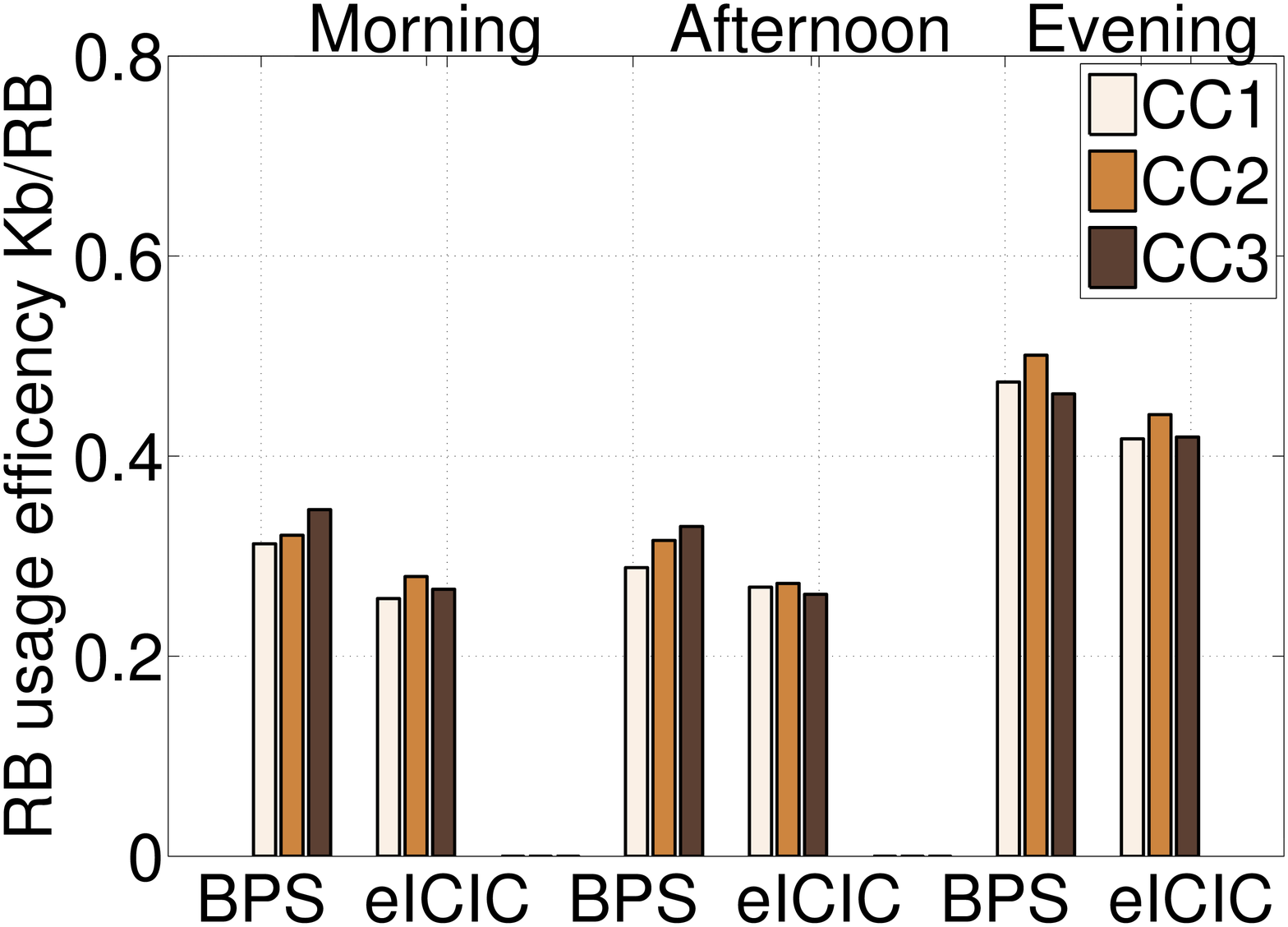}
\hspace{-2mm}
\includegraphics[width=0.33\textwidth]{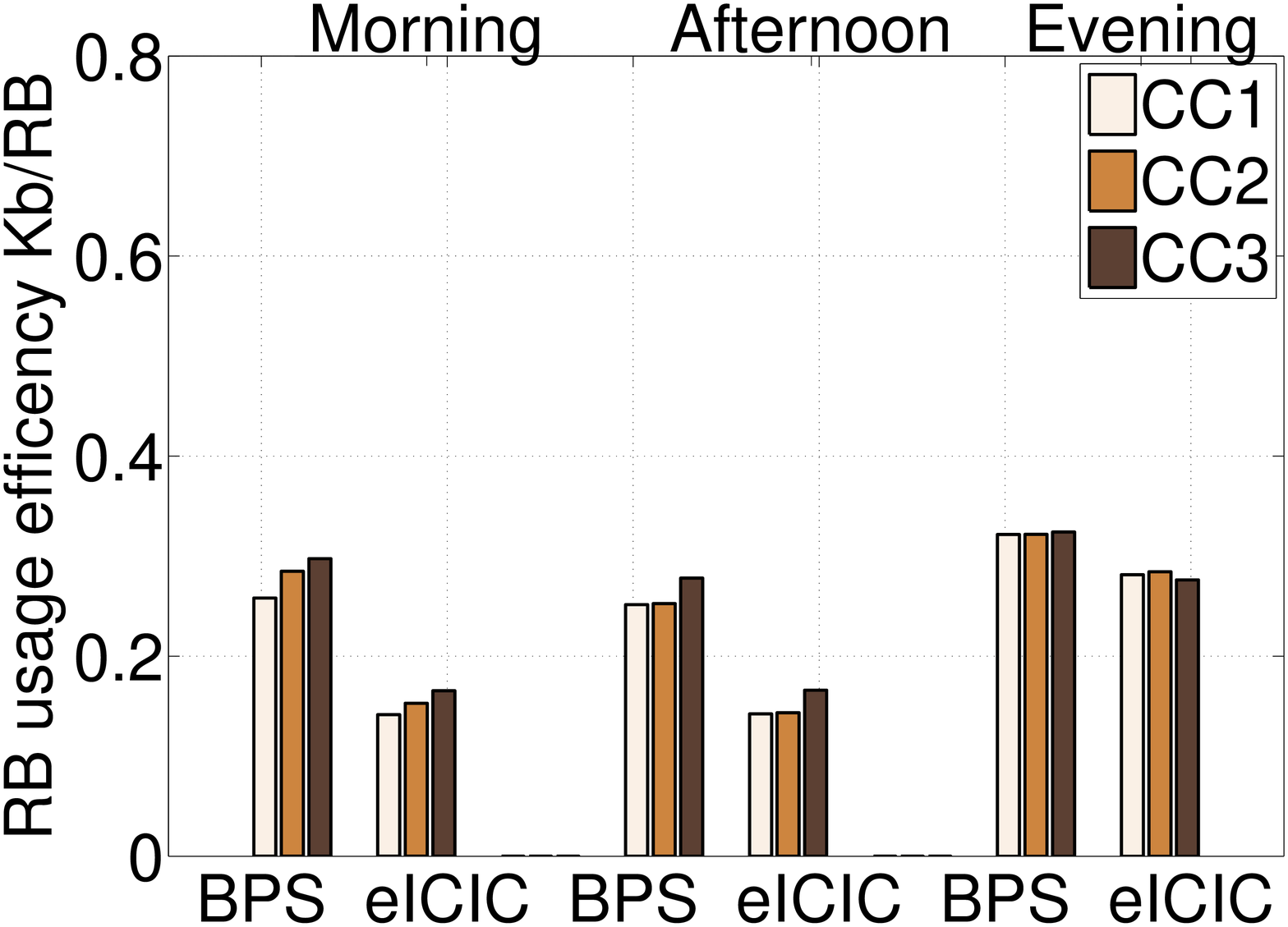}	
\hspace{-1mm}
\includegraphics[width=0.33\textwidth]{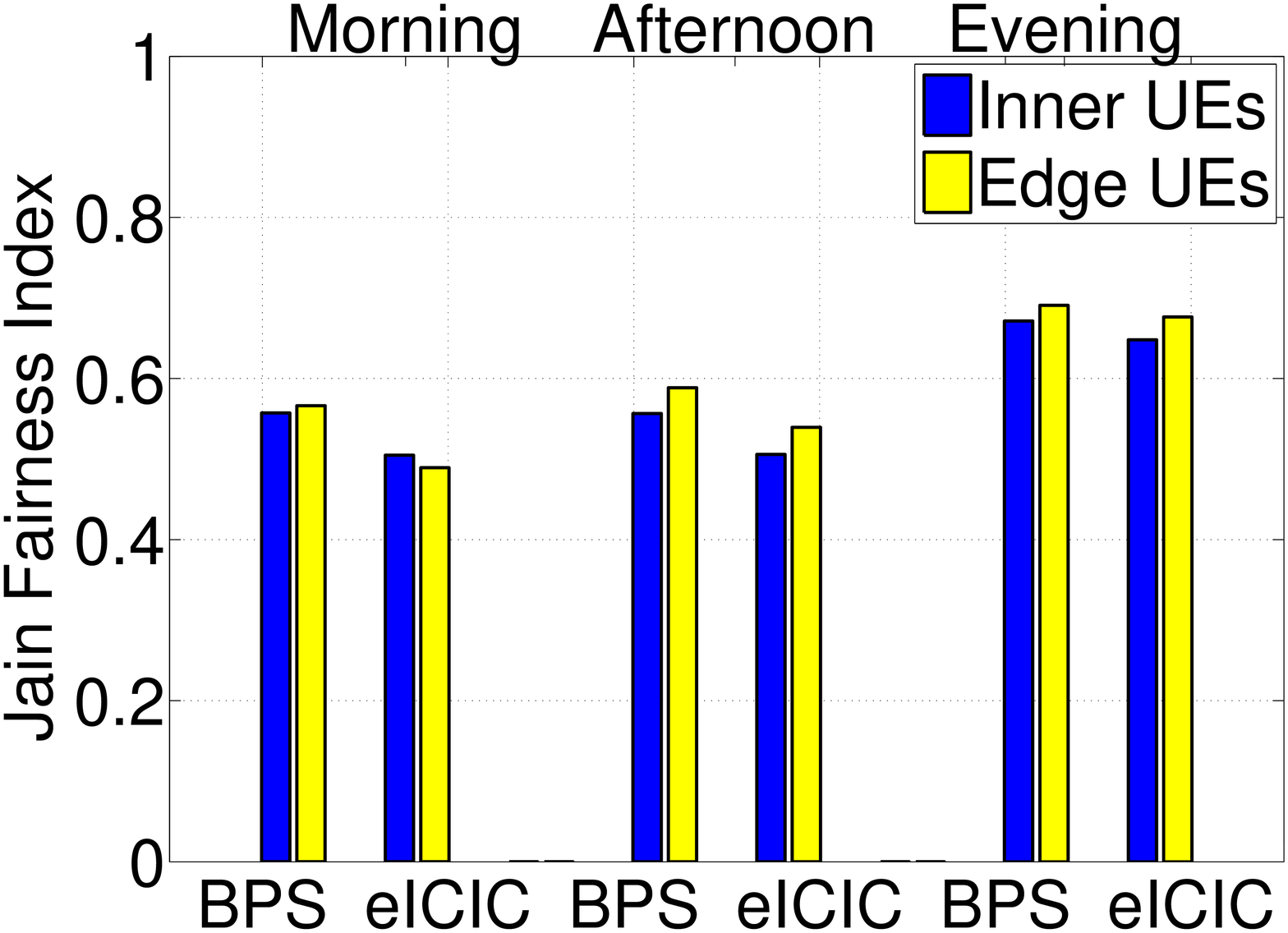}
\vspace{-5mm}
\caption{\label{fig:rbusage} RB usage efficiency expressed in Kb transmitted per RB.  Left: Macro PoAs; Middle: Micro PoAs; Right: Fairness among inner and edge UEs (Jain's index).}\vspace{-5mm}
\end{figure*}

Fig.~\ref{fig:rbusage}  shows the RB usage efficiency for macro (left) and micro PoAs (middle) calculated in terms of kilobits transmitted per number of RBs used. Note that this metric takes into account only those RBs allocated to UEs, not the overall number of RBs available. Again, BPS improves the performance of the network for all types of  PoAs, but more significantly for micro PoAs. eICIC alone introduces important improvement in this metric, especially for macro PoAs, by offloading some of their UEs to the micro PoAs; however, BPS provides an additional edge while lowering the overall power consumption, as seen in the previous figures.  For micro PoAs, BPS improves this metric significantly by strategically varying the  transmit power of the different macro PoAs to reduce the overall interference. The performance of micro PoAs is further improved by the fact that the power setting of the micro PoAs within the same cell is decided at the team level, ensuring optimal coordination in terms of interference. It is worth noting that BPS could also be applied jointly with eICIC, especially to take advantage of the CRE feature. 

\begin{figure*}
\centering
\hspace{-5mm}
\includegraphics[width=0.33\textwidth]{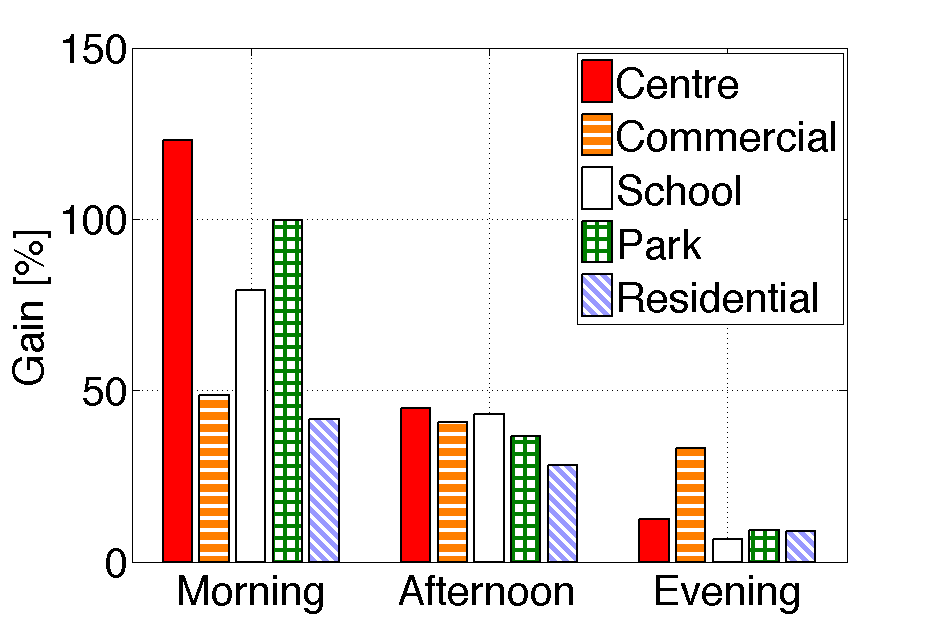}	
\hspace{-5mm}
\includegraphics[width=0.33\textwidth]{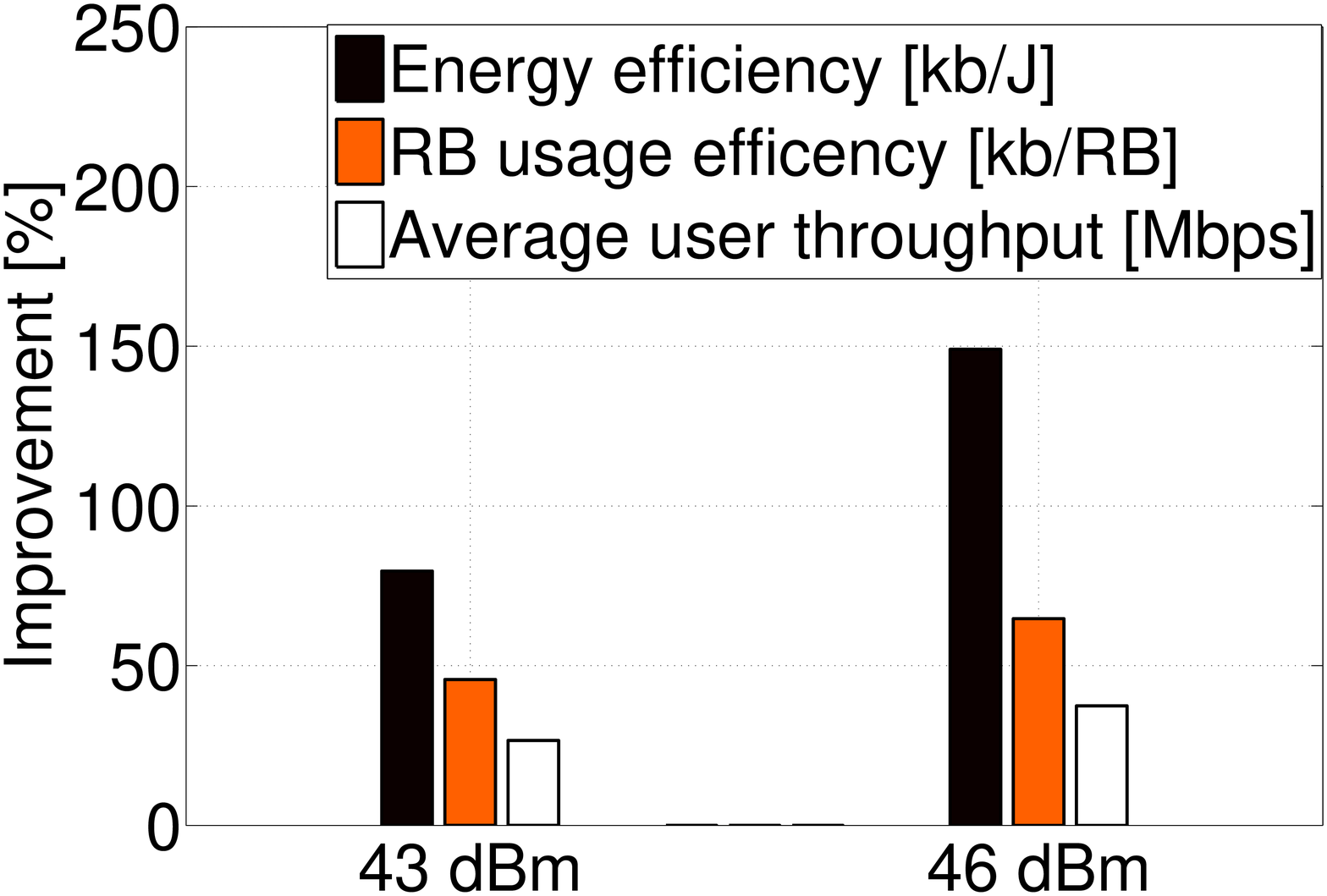}
\hspace{-5mm}
\includegraphics[width=0.33\textwidth]{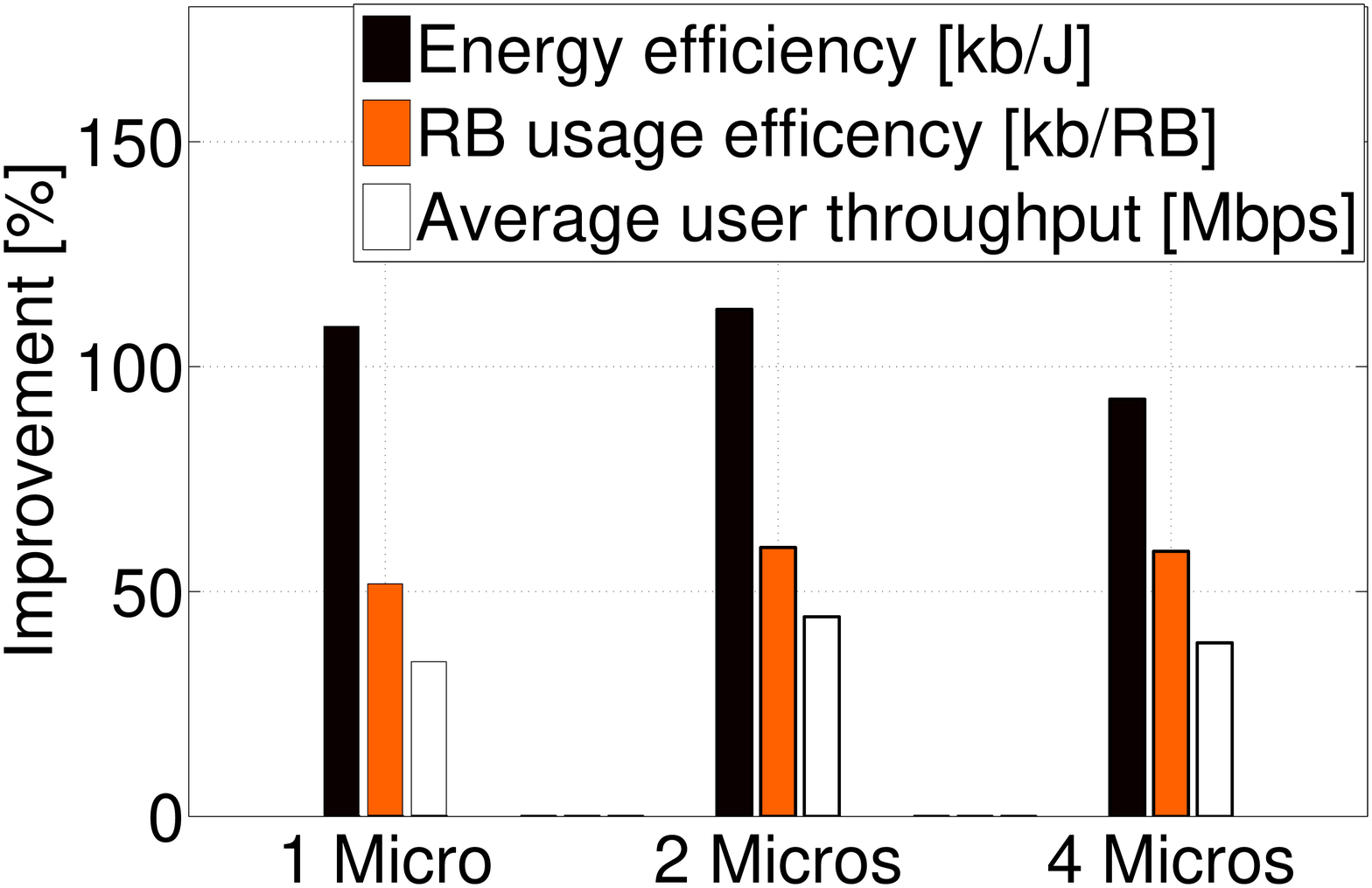}
\vspace{-5mm}
\caption{\label{fig:variations}Left: BPS gains in average user throughput in the different urban areas and at different times of the day. Middle and Right: Improvement due to BPS over eICIC in energy efficiency, RB usage efficiency, and average user throughput,  for a varying number of micro PoAs within a cell (Middle) and  different maximum transmit power for macro PoAs (Right). }\vspace{-5mm}
\end{figure*}

Fig.~\ref{fig:rbusage} (right)  shows the level of fairness between inner and edge UEs in terms of average user throughput, by calculating the Jain fairness index. While the level of fairness for inner and edge UEs tends to be the same, there is a modest improvement for both categories when BPS is applied. Note  that, in multi-tier networks with high density of small cells, the line between inner and edge UEs tends to blur, as edge UEs under the coverage of a micro PoA may experience even better conditions than inner UEs; as a result, the average throughput may vary greatly between UEs of the same category. BPS, however is able to improve the fairness by limiting the overall interference.  Fig.~\ref{fig:variations} (left) depicts the gains obtained by using BPS, compared to eICIC,  in terms of average user throughput in the different urban areas;  again, significant gains are shown, especially during morning and afternoon.

Finally, in Fig.~\ref{fig:variations} (middle), we look at the improvement obtained by applying BPS when compared to eICIC alone, for different network configurations with a varying number of microcells within each cell. The improvement in the three core metrics: energy efficiency, average user throughput and RB usage efficiency, tends to be significant and consistent as the number of microcells is increased. In Fig.~\ref{fig:variations} (right), we also show the gains achieved in the same core metrics, when we consider a higher maximum power for macro PoAs, i.e., 46~dBm (40~W), which is foreseen for 5G systems \cite{metis}, instead of 43~dBm (20~W), which we typically assume. As expected, the effect of BPS is increased when the maximum power of the macro PoAs is elevated, since the effects of interference, which BPS effectively mitigates, are even more pronounced.

\vspace{-5mm}

\vspace{2mm}
\section{Conclusions\label{sec:concl}} 

Given the devastating effects interference will have in future networks as they become more dense and heterogeneous, effective means to contain and mitigate it will be key to enabling the optimal use of resources. In this paper, we proposed a novel solution for downlink power setting in dense networks with carrier aggregation, which aims to reduce interference and power consumption, and to provide high quality of service to users. Our approach leverages the different propagation conditions  of the carriers and the different transmit powers that the various types of PoAs in the network can use for each carrier. 
Applying game theory, we framed the problem as a competitive game among teams of macro and micro PoAs, and identified it as a game of  strategic substitutes/complements with aggregation.  
We then introduced a distributed algorithm that enables the teams to reach a desirable NE  
in very few iterations. Simulation results, obtained in a realistic large-scale scenario, show that our solution greatly outperforms the existing  strategies in the main performance metrics, such as energy efficiency, user throughput and spectral efficiency, while consuming little power. 
At last we remark that, while in this paper we focused on downlink power setting, our approach could be applied to uplink power control as well. In particular, in future  ultra dense cellular networks, the set of users accessing a small cell PoA could be modelled as a team whose leader is the PoA itself. Then the goal would be to set the uplink transmit power so as to mitigate the interference that such users may cause at the neighboring PoAs providing service to other sets of users.

\section*{Acknowledgment}
This work has received funding from the 5G-Crosshaul project (H2020-671598).

\section*{Appendix A: Proof of Theorem 1}
 
Let us first consider the single-carrier case; the interference expression given in Eq.~(\ref{eq:interference}) becomes $I^t(\boldsymbol{s^{-t}}) = \sum_{t'\in\Tc \wedge t'\neq t}s^{t'}a_{t'}$. 
It is clear that this expression fits the aggregator definition provided in Eq.~(\ref{eq:aggregator}), and, consequently, game $G$ meets the conditions set out in  Eqs.~(\ref{eq:cond-1})-(\ref{eq:cond-2}) and in either  Eq.~(\ref{eq:cond-3}) or Eq.~(\ref{eq:cond-4}), as shown in \cite{pa-potential}. The extension to a multi-carrier game with  multi-location teams, implies that the strategy chosen by the team is not a scalar value but a matrix. Likewise,  the interference experienced by each team (i.e., the  aggregator) is a matrix. Without loss of generality, we can assume that the number of locations is the same in each team, and the set of the available carriers is the same for all teams. A decision has to be made for each location and each carrier. As already defined, the strategy of a team $t$, $\boldsymbol{s^t}$, is now an $L\times C$ matrix, while the team interference caused to team $t$, aggregated from other teams' strategies, can now be modelled as a $|Z_t|\times C$ matrix, each element of which is given by Eq.~(\ref{eq:interference}). The $\boldsymbol{I^t}$ matrix can be therefore expressed as:
$\boldsymbol{I^t} = \sum_{t'\neq t}\sum_{l'\in\Lc^{t'}}\boldsymbol{a^{t}_{l'}}\boldsymbol{\sigma^{t'}_{l'}}$, 
where $\boldsymbol{a^{t}_{l'}}$ is a $|Z_t|\times C$ matrix, populated by the attenuation values $a_{l',z,c}$, with each entry $(z,c)$ indicating the attenuation factor from location $l'$ in team $t'$ to tile $z$ in team $t$, on carrier $c$. $\boldsymbol{\bar{\sigma}^{t'}_{l'}}$ is a diagonal $C\times C$ matrix, where $diag(\boldsymbol{\bar{\sigma}^{t'}_{l'}})=[ s^{t'}_{l',c_1} s^{t'}_{l',c_2} \cdots s^{t'}_{l',c_C}]$.
 It is clear that the final interference matrix can be represented as an aggregation of interference matrices caused by each individual team, therefore the aggregator definition still fits. Condition 1) set out in Eq.~ (\ref{eq:cond-1}), is fulfilled due to the very definition of our payoff function, since the dependence on the other teams' strategies is completely captured by the aggregator. 

Regarding conditions 2) and 3), some further explanations have to be made to account for the fact that strategies are no longer single power level (scalar), but instead sets of power levels for the different locations and carriers within the team. Since both interference and strategy are formulated as matrices in our scenario, we have to define what signifies an increase/decrease in interference, and be able to distinguish between higher level strategies/lower level strategies. A natural way to quantify the value of a matrix would be to use the Frobenius norm, in which case, condition 3) becomes:
\vspace{-2mm}
\begin{align}
& ||\theta_t(\boldsymbol{I'^{t}})||_F \leq ||\theta_i(\boldsymbol{I^t})||_F, \,\,\, \forall ||\boldsymbol{I'^t}||_F>||\boldsymbol{I^t}||_F \quad ; \nonumber\\
& ||\theta_t(\boldsymbol{I'^{t}})||_F \leq ||\theta_i(\boldsymbol{I^t})||_F, \,\,\,\forall ||\boldsymbol{I'^t}||_F<||\boldsymbol{I^t}||_F \nonumber
\end{align}
for games of strategic substitutes and games of strategic complements, respectively.

Note that the output of $\theta_t(\boldsymbol{I'^{t}})$ is a strategy, $\boldsymbol{s^t}$, for team $t$ which is, as we said, a matrix. Similarly, we may use the Frobenius norm to differentiate between higher/lower strategies.  To fulfill condition 2), the best-reply function must be continuous, i.e., the output of $\theta_t$ given a specific value of $\boldsymbol{I'^{t}}$ must be unique. In general, there may be cases in which a team may be indifferent between several strategies, in terms of payoff. In such a scenario, we consider that the team can apply the list of preferences (i)-(iii) in Sec. \ref{subsec:game-definition} to fulfill this condition. Concerning condition 3), we closely analyse the payoff function of the team given in Eq.~(\ref{eq:teampayoff}). We note that the team payoff is a sum of individual payoffs obtained at each tile for each carrier. The payoff in each tile is directly linked to the interference value corresponding to that tile. Since we know that at the individual tile condition 3) holds (it is identical to the single-carrier single-location case), then it will hold also at the team level. Namely, when the level of interference experienced by a specific tile increases, increasing thus the value of the Frobenius norm of the interference matrix, then we know that the best reply of the individual location (which serves the specific tile) will be lower/higher depending on whether the game is of strategic substitutes or complements. Lowering/increasing the transmit power at one of the locations, indicates that the Frobenius norm of the strategy matrix will also decrease/increase. 

\section*{Appendix B: Proof of Theorem 2}

Without loss of generality and for ease of presentation, we consider the simplified scenario we  referred to before, with single-location teams and one tile per location. Indeed, 
recall that the team's payoff is given by its utility, represented by the sigmoid function in Eq.~(\ref{eq:team-utility-sigmoid}), discounted by the cost in Eq.~(\ref{eq:fullcost}). From these expressions, it is easy to see that considering multiple per-team locations and tiles just translates into a larger number of additive terms within the summation, hence the behavior of the expressions does not change. 

Allowing for a continuous, rather than discrete strategy set,  a team's best reply response can be derived by evaluating the first derivative of $w^t$ with respect to $\boldsymbol{s^t}$ and solving for zero, i.e., 
 \begin{align}
 &\boldsymbol{s^t_{br}}(I^t) = \arg\max w^t(\boldsymbol{s^t},\boldsymbol{s^{-t}})= - \frac{I^t+N}{\alpha a}\label{eq:bestr} \nonumber \\ 
 &\left[\ln{\left(\frac{\alpha}{2\xi^t(I^t+N)}-1-\sqrt{\left(1-\frac{\alpha}{2\xi^t(I^t+N)}\right)^2-1}\right)}-\alpha\beta \right] 
 \end{align} 
By computing the derivative of Eq.~(\ref{eq:bestr}) with respect to $I^t$, we get:

\begin{equation}
\frac{\partial\boldsymbol{s^t_{br}}}{\partial I^t}=\frac{\beta}{a}\mathord{-}\frac{1}{a\sqrt{\alpha[\alpha \mathord{-} 4\xi^t(I^t+N)]}} \mathord{-}\frac{1}{\alpha a} \ln\left (x  \mathord{-}\sqrt{x^2 \mathord{-} 1}\right)
\label{eq:der-bestr} 
\end{equation}
where $x=\frac{\alpha}{2\xi^t(I^t+N)}-1$. Looking at  this expression, we can see that for low interference values, i.e., as $I^t\to0$, the first two terms of Eq.~(\ref{eq:der-bestr}) tend to $\frac{1}{a}\left(\beta-\frac{1}{\alpha}\right)$  (assuming $N$ is very low). As interference increases, the value of the first term is unchanged while the second term (which is always positive) increases. With regard to the third term, provided that $x\geq1$, which is always true if the condition identified in Eq.~(\ref{eq-priceubound}) is satisfied, the range of values taken by the logarithm is $(0,1]$. In particular, the value of the logarithm  tends to $-\infty$ when $x\to\infty$ (i.e., $I^t$ is very low), and it reaches $0$ when $x\to 1$  (i.e., $I^t$ tends to its maximum as per Eq.~(\ref{eq-priceubound})), indicating that the third term is always non positive.  Thus, when the interference is low, the value of Eq.~(\ref{eq:der-bestr}) is positive (note the negative sign in front of the third term). As interference increases, the value of the derivative reaches $0$ at one point  and then it becomes negative for higher values of interference (note the negative sign in front of the second term). 
In summary,  when interference is low,  a team's best response  increases with interference, then it reaches a {\em stagnation region}  (where Eq.~(\ref{eq:der-bestr}) takes either 0 or very small values),  after which it starts to decrease slowly as interference becomes too high. 
Also, note that the presence of a stagnation region is due to the behaviour of the utility sigmoid function, which, as shown in Eq.~(\ref{eq:team-utility-sigmoid}), increases quickly initially as it reaches its peak and saturates, varying little once the saturation point has been reached.

Let us now focus on the case of three teams, where the transmission of one  of them has the same effect in terms of interference on the other two, and vice versa. We remark that any scenario where there is a dominant interferer can be viewed as a simpler, two-team game. 
Based on the above findings, when the teams start from zero strategy as foreseen in BPS, the game will evolve as follows. The first team to play will increase its power, in order to receive a payoff that is higher than zero. However, since the interference it receives from the other teams is zero, the first team will be satisfied with choosing the leftmost power level that corresponds to the maximum payoff (i.e., the saturation point of the utility sigmoid function). According to the dynamic exhibited by Eq.~(\ref{eq:der-bestr}), the first team's increased transmit power will prompt an increase in the best response of the team who plays next, and so on. 
Such behaviour of the teams will repeat itself similarly in the successive iterations of the game, till the value of interference experienced by the team currently playing falls in the stagnation region. This a turning point in the game, as the current team will find beneficial not to increase its transmit power. 
From this point on, the other teams will not change their strategies, as they will experience the same interference as before, hence they will have no interest to change their transmit power level. This implies that the game has converged to a NE (none of the teams has any incentives to move from its chosen strategy), and,  because each team started from zero, incrementally increasing its strategy (hence the interference), the NE is at the lowest possible strategy level for each team. 

We are now left to show that such an NE is the best in terms of  social welfare, which is defined as the sum of individual payoffs of the teams. 
To this end, let us  
substitute the expression obtained for $\boldsymbol{s^t_{br}}$ (given in Eq.~(\ref{eq:bestr})) in the utility and payoff functions, so that we obtain the expressions that describe the trends of the utility and payoff  when applying the best response, as the interference increases:
\vspace{-2mm}
\begin{align}
u^t_{br}(I^t)&=\frac{2\xi^t(I^t+N)}{\alpha-\sqrt{\alpha^2-4\alpha\xi^t(I^t+N)}} \quad ; \quad \nonumber \\
w^t_{br}(I^t)&=u^t_{br}(I^t)-\xi^t\boldsymbol{s^t_{br}}(I^t) \label{eq:wbr} \,.
\end{align}

From the expressions in~(\ref{eq:wbr}), we note that, as the interference increases, $u^t_{br}$ and $w^t_{br}$ decrease\footnote{Indeed, the derivative of the utility function with respect to the interference is always negative.}, even though the teams are playing their best responses.
Let us assume, by contradiction, that there is a second NE which provides a better social welfare. Because this NE was not reached according to BPS, it must be a point that has not been explored by the algorithm, thus the strategy in this second NE is higher than in the previous one at least for one of the teams.   That is, the overall transmitted power in this second NE must be higher, implying that the teams are facing higher interference than in the previous NE. As shown by the above equations, the utility and payoff values are always decreasing with interference, therefore the utility/payoff values of teams facing higher interference must definitely be lower than those obtained in the previous NE, implying that the social welfare must also be lower. 

At last we remark that the NE reached through our BPS algorithm may not coincide with the global optimal point in terms of social welfare, but it is the optimum among the game's pure NEs. Furthermore, the above proof can be easily extended to any scenario including a generic number of teams, provided that the interference level that each player causes on the others is given.

\end{document}